\documentclass[
print,
superscriptaddress,
 amsmath,amssymb, jmp,
pra,
]{revtex4-2}
\usepackage{gensymb}
\usepackage{graphicx}% Include figure files
\usepackage{dcolumn}% Align table columns on decimal point
\usepackage{bm}% bold math
\usepackage{braket}
\usepackage{mathtools}
\usepackage{amsmath}
\usepackage{amssymb}
\usepackage{appendix}
\usepackage{float}
\usepackage[section]{placeins}
\usepackage{caption}
\usepackage{xcolor}
\usepackage[normalem]{ulem}
\providecommand{\ignore}[1]{}

\newif\ifcmnt
\cmnttrue
\ifdefined\cmntsoff\cmntfalse\fi
\ifcmnt
    \providecommand{\aucmnt}[1]{#1}

\else
    \providecommand{\aucmnt}[1]{}

\fi
\begin{document}

\author{Ariel Shlosberg}
\affiliation{Department of Physics, University of Colorado, Boulder, Colorado, 80309, USA}
\affiliation{JILA,  University  of  Colorado/NIST,  Boulder,  CO,  80309,  USA}
\author{Alex Kwiatkowski}
\affiliation{Department of Physics, University of Colorado, Boulder, Colorado, 80309, USA}
\affiliation{National Institute of Standards and Technology, Boulder, Colorado 80305, USA}
\author{Akira Kyle}
\affiliation{Department of Physics, University of Colorado, Boulder, Colorado, 80309, USA}
\author{Graeme Smith}
\affiliation{Department of Physics, University of Colorado, Boulder, Colorado, 80309, USA}
\affiliation{Institute for Quantum Computing and Department of Applied Mathematics,
University of Waterloo, 200 University Ave W, Waterloo, ON N2L 3G1, Canada}

\begin{abstract}
    Quantum key distribution (QKD) seeks to provide a method of generating cryptographically-secure keys between remote parties while guaranteeing unconditional security. Implementations of high-dimensional QKD using dispersive-optics (DO-QKD) have been proposed to allow for multiple secure bits to be transmitted per photon while remaining cost-effective and scalable using existing telecommunication technology \cite{mower_2013}. In the recent literature, there have been a number of experimental realizations of DO-QKD systems \cite{PhysRevA.90.062331,10.1063/1.5089784,10.1063/5.0002595,10.1088/2058-9565/acfe37,10.1088/2058-9565/ad0f6f}, with security analysis based on the treatment in Ref.~\cite{mower_2013}. Here we demonstrate that in the case of finite dispersion, the model assumed for the eavesdropper's attack in Ref.~\cite{mower_2013} is non-optimal for the eavesdropper, which leads to a significant overestimation of the secure key rate between parties. We consider an alternative attack model that Alice and Bob find indistinguishable from the Ref.~\cite{mower_2013} model, as long as they are restricted to making the measurements typical in DO-QKD. We provide concrete examples where a significant gap exists between the Holevo information, and therefore the secret key rate, predicted by the two models. We further analyze the experiment in Ref.~\cite{PhysRevA.90.062331} as an example of a case where secure key is predicted according to the Ref.~\cite{mower_2013} model, but where in fact there is zero secure key rate when considering the full set of collective attacks that an eavesdropper may perform.
\end{abstract}

\title{Security Assumptions in Dispersive-Optics QKD}
\maketitle
\section{Introduction}
Quantum key-distribution (QKD) is a scheme that leverages the properties of quantum mechanics to generate cryptographically-secure keys between remote parties \cite{BENNETT20147,PhysRevLett.67.661}. The original implementations of QKD, such as BB84, focused on transmitting qubits, allowing up to a maximum of a single bit of shared key to be generated per measurement \cite{BENNETT20147}. In order to increase key rates beyond what is obtainable through the transmission of optical qubits, high-dimensional (HD) or continuous-variable (CV) physical degrees of freedom can be used to distribute more than one secure bit per detected photon \cite{Etcheverry2013,mower_2013,PhysRevLett.96.090501,PhysRevLett.112.120506,Nunn:13,PhysRevLett.88.057902,PhysRevA.63.052311,PhysRevLett.93.170504}. There are many choices for which degrees of freedom to use to transmit information, including the commonly employed position-momentum and time-frequency bases \cite{Grosshans2003,PhysRevLett.100.110504,Qi:06,PhysRevA.105.052429}. Generally, QKD protocols provide security by having Alice and Bob make measurements in conjugate or mutually unbiased bases, whether in the discrete or continuous-variable settings \cite{raul_thesis}. Most QKD protocols can be categorized as either prepare-and-measure schemes, where Alice prepares a random quantum state from some ensemble and Bob performs a measurement from a pre-established set, or as entanglement-based protocols, where entanglement is used as a resource to generate the randomness and security needed for QKD. However, the family of prepare-and-measure schemes can be unified under an entanglement-based representation since the purification of the ensemble that Bob receives is itself an entangled quantum state
\cite{PhysRevLett.68.557,raul_thesis,10.5555/2011564.2011570}. 

There are multiple different proposals and implementations for entanglement-based CV-QKD, but in this work we will focus on the set of dispersive-optics (DO-QKD) protocols, which involve an entangled state produced by a nonlinear crystal through spontaneous parametric down-conversion (SPDC) \cite{PhysRev.124.1646, PhysRevLett.25.84}, and time-tagged single-photon measurements.
In general, asymptotic security is based on the assumption that Alice and Bob know the initial entangled pure state and make trusted measurements in order to constrain any effect that an eavesdropper, Eve, can have on the state. 
Assuming Eve has access to the purification of the final state, the information that Eve can access is constrained by the Holevo information between Eve's subsystem and Alice's measurement outcomes.
In DO-QKD, two incompatible basis measurements are made; the first are direct arrival-time measurements and the second involve applying a dispersive optical element into the beam path before a timing measurement. The security analysis of many DO-QKD experiments \cite{PhysRevA.90.062331,10.1063/1.5089784,10.1063/5.0002595,10.1088/2058-9565/acfe37,10.1088/2058-9565/ad0f6f} is based on the treatment in Ref.~\cite{mower_2013}, which provides a procedure to calculate the secure key rate based on experimental measurements.
Using the changes in correlation times between Alice's and Bob's timing measurements (with and without applying dispersion), they are able to place bounds on the form and intensity of Eve's attacks. 
The treatment in Ref.~\cite{mower_2013} considers a particular variant for Eve's attacks, which is inspired by the class of phase-insensitive channels acting on the state in the limit of infinitely large dispersion \cite{e18010020, PhysRevResearch.2.013208}. 
In this high-dispersion limit, the state that Alice and Bob share (the DO-QKD state) becomes a two-mode squeezed vacuum  (TMSV) state.
In principle, Alice and Bob may share any physical channel, and in many QKD settings, it is assumed that they share an experimentally-motivated thermal, lossy channel. 
However, to guarantee unconditional security against all collective attacks, it is crucial to make measurements that constrain the channel between Alice and Bob, or alternatively the final state that they share, to be of the assumed form.   
Prior work in Ref. \cite{PhysRevLett.112.120506} has argued that for a TMSV state, the attack in Ref.~\cite{mower_2013} is optimal for Eve. By optimal, we mean that Eve chooses an attack that minimizes the secret key rate between Alice and Bob subject to their experimental observations.
However, the attack in Ref. \cite{mower_2013} is not sufficiently general to capture the full range of collective attacks that Eve may perform for finite dispersion. 
In fact, multiple papers and experiments have used the form for Eve's attack given in Ref. \cite{mower_2013}, implicitly assuming a phase-insensitive channel acts on a TMSV state (see Appendix \ref{app:physical_mower}), while in practice Alice and Bob make measurements on a joint  state that involves local squeezing \cite{e18010020}. 
As a result, the security analysis of several experiments relies on an assumption that is not met in practice, as most experiments operate well below the regime where the TMSV state approximation may be reasonable \cite{PhysRevA.90.062331,10.1063/1.5089784,10.1063/5.0002595,10.1088/2058-9565/acfe37,10.1088/2058-9565/ad0f6f}. This leads to a theoretical flaw in the security analysis of these experiments. There have also been multiple numerical analyses that appear to have not fully considered the role that weak dispersion can have on the security model \cite{PhysRevA.91.022336,LIU2021100119}.

In this work, we provide a concise description of this security issue and demonstrate that it can lead to a significant impact on the secure key rate of practical DO-QKD experiments. 
We illustrate this point with numerical examples using parameters that correspond to recent experiments. This paper is organized as follows. In Section \ref{sec_prelim} we provide an overview of the relevant aspects of DO-QKD and Gaussian quantum information. We lay out the Ref. \cite{mower_2013} model and explain the proposed process for bounding the information leaked to Eve.
%We describe how the security analysis of many DO-QKD papers implicitly makes the assumption that the initial state shared by Alice and Bob is a TMSV state when constructing the form of Eve's attack. 
In Section \ref{sec_not_epr}, we show that the initial state shared by Alice and Bob is a function of the amount of dispersion applied, and for finite dispersion, the initial state is not TMSV. 
In Section \ref{sec_security_flaw}, we demonstrate numerically that for a representative set of experimental parameters, the discrepancy between the initial state and a TMSV state is significant enough that analyzing according to Ref.~\cite{mower_2013} leads to a significant overestimate of the amount of secret key that is generated. 
In Section \ref{sec_comparisons}, we make numerical comparisons to a previously-published, representative DO-QKD experiment, and provide evidence that this experiment overestimates the amount of secret key that is generated. 
Finally, in Section \ref{sec:proposed_solutions}, we discuss potential strategies to realizing secure key generation using the DO-QKD protocol and the associated limitations that may exist.

\section{Preliminaries}
\label{sec_prelim}
In a general entanglement-based CV-QKD scheme, Alice prepares an entangled system $\rho$, sends part of the state to Bob over a non-secure quantum channel, and then they proceed to generate secret key through LOCC operations, including classical post-processing (error-correction and privacy amplification). 
The specifics of the quantum channel dictate the amount of information that Eve can possibly access about the shared key, as we give Eve the purification of the final state.
While Alice and Bob may share an arbitrary quantum state, the secret key rate can be lower bounded by assuming that Alice and Bob instead share a Gaussian state with equivalent first and second moments ($\mathbf{\bar{x}},\ \mathbf{\Gamma}$) \cite{PhysRevLett.96.080502}. The bosonic state can be described by quadrature field operators $\mathbf{\hat{x}}=(\hat{x}_1,\hat{p}_1,\dots,\hat{x}_N,\hat{p}_N)$, and the first and second moments are given by $\mathbf{\bar{x}}=\text{Tr}(\rho\mathbf{\hat{x}})$ and $\mathbf{\Gamma}_{ij}=\frac{1}{2}\text{Tr}(\{\hat{x}_i-\bar{x}_i,\hat{x}_j-\bar{x}_j\}\rho)$, where the anti-commutator notation means $\{\hat{x}_i,\hat{x}_j\}=\hat{x}_i\hat{x}_j+\hat{x}_j\hat{x}_i$ \cite{RevModPhys.84.621}.
The covariance matrix must obey the uncertainty principle $\mathbf{\Gamma}+\frac{i}{2}\mathbf{\Omega}\geq0$ with the symplectic form $\mathbf{\Omega} =\bigoplus_{i=1}^n \bigl(\begin{smallmatrix} 0 & 1\\ -1 & 0
\end{smallmatrix}\bigr)$.
The two-mode Gaussian covariance matrix can be written as
\begin{equation*}
     \Gamma= \begin{pmatrix}
    \gamma_{AA} & \gamma_{AB}\\
    \gamma_{AB}^T & \gamma_{BB}
    \end{pmatrix},\quad \gamma_{AB} = \frac{1}{2}\begin{pmatrix}
        \langle\{\hat{x}_A,\hat{x}_B\}\rangle, \langle\{\hat{x}_A,\hat{p}_B\}\rangle\\
        \langle\{\hat{p}_A,\hat{x}_B\}\rangle, \langle\{\hat{p}_A,\hat{p}_B\}\rangle
    \end{pmatrix}.
\end{equation*}

Gaussian attacks, which map Gaussian states to Gaussian states, have been proven optimal in cases where measurements of the covariance matrix are used to constrain the final state. \cite{Cerf_gaussian_optimality,PhysRevLett.97.190502,PhysRevA.81.062314}. 
To guarantee unconditional security, the most general Gaussian attack that Eve can implement is a coherent attack, where Eve interacts a global ancillary system with all of the signals that Alice sends to Bob \cite{RevModPhys.84.621}. 
Eve stores this global state until after intercepting the classical communication between Alice and Bob, at which point she chooses a joint measurement to perform. 
Alternatively, Eve can perform a collective attack, which is defined by her interacting independently with each system that Alice transmits to Bob, generating a state of the form $\rho_{ABE}^{\otimes n}$.
Since the QKD scheme considered here is permutation invariant with respect to the order that Alice and Bob communicate the individual states, then considering the set of collective attacks is sufficient to guarantee security in the asymptotic regime \cite{PhysRevLett.102.110504}. 
A general Gaussian collective attack can be modelled by allowing any one-mode Gaussian channel on Bob's system and giving Eve the purification of the joint output state \cite{PhysRevLett.101.200504}. 
Following the standard model of QKD, we assume that Alice begins by preparing a pure state, $\rho_{AB}$, and then proceeds to send  part of the state to Bob over a noisy channel, at which point Alice and Bob end up sharing the mixed state $\rho_{AB}'$. To determine how much information Eve can obtain in this setup, we consider the purification, $\psi_{ABE}$, of the mixed state, where Eve holds $\rho_E=\text{Tr}_{AB}(\psi_{ABE})$.
The secure key rate is given as $\Delta K=\min_{\mathbf{\Gamma}'\in \mathcal{M}} [I(A:B)-\chi(A:E)]$, where $\mathcal{M}$ is the set of all physical covariance matrices $\Gamma_{AB}'$ that are consistent with experimental observations, $I(A:B)$ is the mutual information between Alice's classical outcomes and Bob's classical outcomes, and $\chi(A:E)$ is the Holevo information between Eve's subsystem and Alice's classical outcomes \cite{PhysRevA.72.012332,Devetak_2005}. Mutual information is defined in the standard way: $I(A:B) = H(A)-H(A|B)$, where $H(X)$ is the Shannon entropy of the classical random-variable $X$. The Holevo information is given by $\chi(A:E)=S(E)-\sum_x p(x) S(E|X_A)$, where $S$ is the von-Neumann entropy and $X_A$ is a random variable that denotes Alice's measurement outcomes \cite{cerf1996accessible}. 

The set $\mathcal{M}$ is determined by the type (and precision) of measurements that Alice and Bob make on their respective systems, and in the usual DO-QKD setting, the set is primarily constrained by all correlated measurements performed (concrete list provided further in text). If instead, Alice and Bob perform tomography of the second moments of the state and obtain a particular covariance matrix, $V$, then $\mathcal{M}$ becomes the set containing only $V$ and the measured covariance matrix directly determines the secure key rate.
In DO-QKD, Alice and Bob measure the changes in correlated variances using a variety of basis combinations, which can be written as sums of elements of the covariance matrix. Each of the measured correlated variances defines a relation between entries of the final covariance matrix and experimental outcomes, which Alice and Bob use to constrain Eve's effect on the state. However, if one does not consider the full set of matrices consistent with observations, rather only considering a restricted set (such as done in Ref.~\cite{mower_2013}), then it can lead to a significant overestimation of the secure key rate.
% However, if Alice and Bob start with a more general state this simplification is not accurate (reword this later).
% As we will show, the security analysis of several DO-QKD experiments (citations) assumes that the one-mode channel is of this form, even when the initial state is not a TMSV pair. This leads to an overestimation of the secure key that can be significant in numerical examples that we consider(reword this later).

In the setting of DO-QKD, we assume that the biphoton state generated by a pumped SPDC source can be expressed as
\begin{equation}
    \label{eq:DO-QKD_state}
    \ket{\Psi_{AB}} = \int\int dt_a dt_b \frac{1}{\sqrt{2\pi\sigma_{coh}\sigma_{cor}}} e^{-(t_a +t_b)^2/(16\sigma_{coh}^2) - (t_a-t_b)^2/(4\sigma_{cor}^2) - 
   i \frac{\omega_p}{2}(t_a+t_b)} \ket{t_at_b},
\end{equation}
where $\sigma_{coh}$ and $\sigma_{cor}$ are time constants set by the laser coherence time and the phase-matching bandwith of the SPDC source, and $\omega_p$ is the laser pump frequency \cite{mower_2013, PhysRevLett.98.060503}. 
The conjugate bases (or quadratures) that we use for measurement in this QKD setting correspond to the arrival time of the photons, $\hat{T}_j=\int t_j\ket{t_j}\bra{t_j}dt_j$, and the dispersed-arrival time, $\hat{D_j}=\frac{2}{k_j}\hat{U}_j^\dagger\hat{T}_j\hat{U}_j$, where $j$ denotes the system (A/B) and $\hat{U}_j=\frac{1}{\sqrt{\pi|k_j|}}\int\int e^{-i(t_1-t_2)^2/k_j}\ket{t_1}\bra{t_2}_jdt_2dt_1$. The parameter $k$ determines the strength of the dispersion applied to the optical beam path when making $\hat{D}$ measurements. The choice of normalization for the dispersed-arrival time operator was made in order to satisfy the canonical commutation relation, $[\hat{T}_j,\hat{D}_k]=i\delta_{jk}$. 
The dispersed-arrival time is supposed to provide information about the frequency of measured photons, but as will be shown, is not an exact stand-in. Alice and Bob apply opposite dispersion of equal magnitude ($k_1 = -k_2$) in order to retrieve the temporal correlations after application of group velocity dispersion in the large coherence time limit \cite{PhysRevA.45.3126}. Regardless of the measurement basis, Alice and Bob make timing measurements with their detectors, but when in the dispersed-arrival time basis they measure the operator $\hat{U}_j^\dagger\hat{T}_j\hat{U}_j=\frac{k_j}{2}\hat{D}_j$.

The attack model ascribed to Eve in Ref.~\cite{mower_2013} is defined by the following transformation, 
\begin{equation}
    \Gamma \rightarrow \Gamma'=\begin{pmatrix}
        \gamma_{AA} & (1-\eta)\gamma_{AB} \\
        (1-\eta)\gamma_{BA} &(1+\epsilon)\gamma_{BB}
    \end{pmatrix},
    \label{eq:noise_cov}
\end{equation}
where $\Gamma'$ denotes the covariance matrix of the state after Eve's interference.
In this model, Ref.~\cite{mower_2013} allows $\epsilon$ and $\eta$ to take any values that correspond to physical covariance matrices. As previously stated, the motivation for this attack model is that it may be optimal in the high-dispersion limit, where the SPDC biphoton state becomes a TMSV state \cite{PhysRevLett.112.120506}. 
In the experiments and theory proposals for DO-QKD, Alice and Bob make correlated measurements to put constraints on the covariance matrix, and therefore, the Holevo information that is accessible to Eve. 
They constrain themselves to only making measurements of the correlated variances since direct measurements of individual entries in the covariance matrix are either challenging, or outright impossible, to perform. The different basis combinations that Alice and Bob measure in order to ensure security against Eve are the following: $\{\text{Var}(\hat{T}_A-\hat{T}_B), \text{Var}(\frac{k}{2}\hat{D}_A+\frac{k}{2}\hat{D}_B)$, $\text{Var}(\hat{T}_A+\frac{k}{2}\hat{D}_B)$, $\text{Var}(\hat{T}_B-\frac{k}{2}\hat{D}_A)\}$, where $\text{Var}(\hat{O}) = \text{Tr}(\rho \hat{O}^2)-\text{Tr}(\rho \hat{O})^2$ \cite{mower_2013,PhysRevA.90.062331}.
Throughout the rest of this paper, we refer to this set of measurement results as the correlated variances.

We will show in Sec~\ref{sec_security_flaw} that it is possible to construct physical one-mode Gaussian channels that have identical correlated variances while giving Eve more information than predicted by the Ref.~\cite{mower_2013} model. There they introduce the parameter $\xi_1: \text{Var}(T_A'-T_B') = (1+\xi_1)\text{Var}(T_A-T_B)$, where primed variables correspond to values after Eve's interference, and the following relation can be derived,
\begin{equation}
\epsilon = \frac{4 \xi_1 \sigma_{cor}^2 - \eta \left(8 \sigma_{coh}^2 - 
    2 \sigma_{cor}^2\right)}{4 \sigma_{coh}^2 + \sigma_{cor}^2}.
% \epsilon = \frac{-2\eta\left(d^2-\frac{1}{4}\right)+\xi_1}{\left(d^2+\frac{1}{4}\right)}.
\label{eq:relation}
\end{equation}
There is an equivalent relationship defined by $\text{Var}(\frac{k}{2}D_A'+\frac{k}{2}D_B') = (1+\xi_2)\text{Var}(\frac{k}{2}D_A+\frac{k}{2}D_B)$. To simplify the analysis, we will assume for the rest of this work that the relative changes measured in the arrival time and dispersed-arrival time bases are equivalent (i.e. $\xi_1=\xi_2=\xi$). 
In order to maintain a physical covariance matrix, specifically satisfying the condition $V\geq 0$, we require that $\eta\lessapprox\sqrt{1+\xi}\frac{\sigma_{cor}}{\sigma_{coh}}$. Using the relation in Eq.~\ref{eq:relation}, we can derive the full set of changes in the correlated variances, resulting from Eve's attack (given in Appendix \ref{app:correlated_variances}). 
As we will show in the next sections, if the initial state is not a TMSV state, then the attack model in Eq.~\ref{eq:noise_cov} is not optimal for Eve. This can lead to an overestimation of the secret key shared by Alice and Bob. We provide evidence that existing experiments and results in the literature indeed overestimate the secret key rate in their analyses.

\section{DO-QKD State is not a TMSV State}
\label{sec_not_epr}
The covariance matrix, $\Gamma$, of the Gaussian state in terms of the arrival time and dispersed-arrival time quadratures takes the same form as in Ref.~\cite{mower_2013}, up to a relabeling of Alice's and Bob's systems. We define the unitless quadratures $\widetilde{x}_i = \frac{1}{\sqrt{2\sigma_{coh}\sigma_{cor}}}\hat{T}_i$ and $\widetilde{p}_i=\sqrt{2\sigma_{coh}\sigma_{cor}}\hat{D}_i$ to obtain a unitless form for the matrix:
\begin{gather}
\renewcommand*{\arraystretch}{2}
    \Gamma= \begin{pmatrix}
    \gamma_{AA} & \gamma_{AB}\\
    \gamma_{AB}^T & \gamma_{BB}
    \end{pmatrix} \\
    \gamma_{AA} = \begin{pmatrix}
        \frac{u+v}{4\sqrt{uv}} & \frac{u+v}{8k} \\
        \frac{u+v}{8k} & \frac{(u+v)(4k^2+u v)}{16k^2 \sqrt{uv}}
    \end{pmatrix},
    \gamma_{AB}=\begin{pmatrix}
        \frac{u-v}{4\sqrt{uv}} & -\frac{u-v}{8k} \\
        \frac{u-v}{8k} & -\frac{(u-v)(4k^2+u v)}{16k^2 \sqrt{u v}}
    \end{pmatrix},
    \gamma_{BB} = \begin{pmatrix}
        \frac{u+v}{4\sqrt{uv}} & -\frac{u+v}{8k} \\
        -\frac{u+v}{8k} & \frac{(u+v)(4k^2+u v)}{16k^2\sqrt{uv}}
    \end{pmatrix},
    \label{eq:cov_noiseless}
\end{gather}
where $u=16\sigma_{coh}^2$ and $v=4\sigma_{cor}^2$. The covariance matrix given in Equation \ref{eq:cov_noiseless} is local-unitary equivalent to a two-mode squeezed state. The corresponding symplectic transform that takes the DO-QKD state given by Eq. \ref{eq:DO-QKD_state}, in the unitless ($\widetilde{x},\widetilde{p}$) basis, to the TMSV state is given by 
\begin{gather*}
    S = S_1\oplus S_2 \\
    S_1 = \begin{pmatrix}
    (1+\alpha)^\frac{1}{2} & -(1+\alpha^{-1})^{-\frac{1}{2}}\\
        0 &(1+\alpha)^{-\frac{1}{2}}
    \end{pmatrix},\quad
    S_2 = \begin{pmatrix}
    (1+\alpha)^\frac{1}{2} & (1+\alpha^{-1})^{-\frac{1}{2}}\\
        0 &(1+\alpha)^{-\frac{1}{2}}
    \end{pmatrix},
\end{gather*}
where $\alpha\equiv \frac{uv}{4k^2}=\frac{16}{k^2}\sigma_{coh}^2\sigma_{cor}^2$.
The transform $S$ obeys the symplectic condition, $S\Omega S^T = \Omega$, and diagonalizes the blocks of the covariance matrix:
\begin{gather*}
    \tilde{\Gamma} = S\Gamma S^T = \begin{pmatrix}
        \beta \mathbf{I} & \sqrt{\beta^2-1}\mathbf{Z} \\
        \sqrt{\beta^2-1}\mathbf{Z} & \beta \mathbf{I}
    \end{pmatrix},
\end{gather*}
where $\beta = \frac{\sigma_{coh}}{2 \sigma_{cor}} + \frac{\sigma_{cor}}{8\sigma_{coh}}$. The eigenvalues of $S_1$ and $S_2$ are $e_1 = (1+\alpha)^{\frac{1}{2}}\neq 1$ and $e_2=e_1^{-1}$, which means that the SPDC state requires local squeezing to transform to a TMSV state \cite{RevModPhys.84.621}. When $k<<4\sigma_{coh}\sigma_{cor}$, the DO-QKD state requires a large amount of local squeezing to approach the TMSV state, particularly the squeezing parameter is given by $r=\frac{1}{2}\log(1+\alpha)$. If however, we consider working in the large dispersion limit ($k>>4\sigma_{coh}\sigma_{cor}$), then the unitless covariance matrix goes to $\tilde{\Gamma}$ and the local unitaries approach the identity ($S_1,S_2\rightarrow \mathbf{I}$). The main technical point is that the state considered in these experiments is qualitatively different than a TMSV state. This implies that any security proofs based on symmetries of the TMSV state, and of the measurement protocol, do not directly apply in this setting.

\section{Example of Security Flaw}
\label{sec_security_flaw}
Since we have shown that the DO-QKD state is not a TMSV state, then we expect that the optimal attack for Eve may not simply correspond to a phase-insensitive single-mode channel.
However, it still remains to show that the difference in attack models can lead to a significant gap between the secure key rate that has been predicted in a variety of previous works and the correct secure key rate, subject to general collective attacks. We give a concrete example of such a gap by using the DO-QKD state parameters from Ref.~\cite{mower_2013} and an experimentally realistic value of $k$ from Ref.~\cite{PhysRevA.90.062331}.
We will use the following conventions for Gaussian channels.
An arbitrary Gaussian channel acting on a Gaussian quantum state takes the following form: 
\begin{equation*}
    \mathbf{\bar{x}} \rightarrow \mathbf{T}\mathbf{\bar{x}}+\mathbf{d}, \quad \mathbf{V}\rightarrow\mathbf{TVT^T+N}
\end{equation*}
where $\mathbf{T}$ and $\mathbf{N^T=N}$ are $2N\times 2N$ real matrices and must satisfy the following positivity condition \cite{RevModPhys.84.621}: 
\begin{equation*}
    \mathbf{N}+\frac{i}{2}\mathbf{\Omega}-\frac{i}{2}\mathbf{T\Omega T^T}\geq0.
\end{equation*}

We will consider a specific model for the attack that Eve chooses to implement, $\Phi_0$: a one-mode channel on Bob's system with parameters $\mathbf{T_0}=(1-\eta)\mathbf{I}$ and $\mathbf{N_0}=\frac{1}{2}\text{diag}(\epsilon_1,\epsilon_2)+(\eta-\frac{\eta^2}{2})\mathbf{I}$. 
We note that the parameter $\eta$ is not the standard attenuation parameter, but we make this choice for consistency with Ref.~\cite{mower_2013}.
This channel, which consists of loss followed by the addition of phase-sensitive noise, satisfies the positivity condition as long as $\epsilon_1,\epsilon_2\geq 0$ and $0\leq\eta\leq1$. 
The experimental measurements of the correlated variances, $\text{Var}(T_A'-T_B')=(1+\xi)\text{Var}(T_A-T_B)$ and $\text{Var}\left(\frac{k}{2}D_A'+\frac{k}{2}D_B'\right)=(1+\xi)\text{Var}\left(\frac{k}{2}D_A+\frac{k}{2}D_B\right)$, give the following constraint equations:
\begin{gather*}
\epsilon_1 = \frac{\sigma_{cor}}{\sigma_{coh}}\xi + \eta \left(\frac{\sigma_{cor}}{\sigma_{coh}}-2\right) + \eta^2 \left(1 - 
\frac{\sigma_{coh}}{\sigma_{cor}} - \frac{\sigma_{cor}}{
    4 \sigma_{coh}}\right) \\
\epsilon_2 = \left(\frac{\sigma_{cor}}{\sigma_{coh}} + \frac{
 16 \sigma_{coh} \sigma_{cor}^3}{k^2}\right)\xi + \eta \left(\frac{\sigma_{cor}}{\sigma_{coh}} + \frac{
    16 \sigma_{coh} \sigma_{cor}^3}{k^2}-2\right) + \eta^2 \left(1 - \frac{\sigma_{coh}}{\
\sigma_{cor}} - \frac{\sigma_{cor}}{4 \sigma_{coh}} -
    \frac{16 \sigma_{coh}^3 \sigma_{cor}}{k^2} - 4 \frac{\sigma_{coh} \sigma_{cor}^3}{
    k^2}\right).
\end{gather*}
We have intentionally chosen an alternate model construction where the results of all correlated variance measurements cannot be used to distinguish between the two models in the lossy channel regime ($0\leq\eta\leq1$). To be more precise, the changes in the correlated variances, as a function of $\xi$ and $\eta$, are identical in the Ref.~\cite{mower_2013} model and our alternate model, laid out in Appendix \ref{app:correlated_variances}. Since the secure key is found by optimizing over the set of all physically consistent covariance matrices, then the key rate that Alice and Bob can generate subject to collective attacks is bounded above by the key rate calculated using the alternate model.
A similar situation ensues for other ranges of the parameter $\eta$, which involves constructing a slightly different alternate channel, although we do not address the details here.

Based on experimental outcomes for a given realization of DO-QKD, we can determine the values that these two distinct models associate to the secure key rate. 
Eve's Holevo information is given by $\chi(A:E)=S(\rho_E)-\frac{1}{2}(S(\rho_{E|T})+S(\rho_{E|D}))$, where we have made use of the fact that Bob's and Eve's conditional covariance matrices are independent of Alice's measurement outcome due to the state being Gaussian. We can make further simplifications by realizing that the overall state is pure, Eve is given the purification of Alice and Bob's state, and that after Alice performs a homodyne measurement, the conditional state is still pure. 
The maximum achievable secure key rate is then given by $\Delta K=I(A:B)-\chi(A:E)$. The mutual information between Alice and Bob can just be evaluated on the relevant $\Gamma'$ after Eve's interference. The classical mutual information for a Gaussian state can be evaluated from the correlation coefficient ($r$):
\begin{equation*}
    I=-\frac{1}{2}\log_2(1-r^2),\quad r = \frac{\text{Cov}(X,Y)}{\sqrt{\text{Var}(X)\text{Var}(Y)}},
\end{equation*}
where $X$ and $Y$ are the random variables corresponding to Alice's and Bob's measurement outcomes respectively \cite{gel'fand1959calculation}.

As an example, we take $\sigma_{coh} = 64\times 30$ ps, $\sigma_{cor} = 30$ ps, $\xi=0.78$, and $\eta = 6.3\times 10^{-5}$ (same numbers as in Ref.~\cite{mower_2013}) and we temporarily assume that there are no sources of noise other than Eve's interference. As a representative example, we take the dispersion coefficient to be $k=0.0039$ $\text{ns}^2$ (used in the experiment in Ref.~\cite{PhysRevA.90.062331}), which corresponds to a group delay slope of around $1500\ \text{ps}/\text{nm}$. In the case of the Ref.~\cite{mower_2013} model, the secure key rate for these parameters is $\text{SKR} = 5.58 - 0.69 = 4.89\ \text{bpc}$ (bits per post-selected coincidence), whereas using the alternate channel, $\Phi_0$, we obtain $\text{SKR} = 5.58 - 5.79 < 0\ \text{bpc}$. From this very straightforward example, we can see that using the Ref. \cite{mower_2013} model, in the case of finite dispersion, leads to a large overestimate of the secure key rate. While there is actually no secure key being generated, the Ref.~\cite{mower_2013} model predicts a key rate of nearly $5$ bpc. 
This problem does not resolve even in cases of larger group delay slopes, such as $|D| = 10000\text{ ps}/\text{nm}$ where the secure key rate is $\text{SKR} = 5.58 - 3.67 = 1.91\ \text{bpc}$, but does become less pronounced. 
The convergence of the rates from the two models is expected because as $k \rightarrow \infty$, then the attack model in Ref.~\cite{mower_2013} captures the alternate model as a special case, subject to the constraint that $\xi_1=\xi_2$.

\section{Numerical Comparison to Previous Experiments}
\label{sec_comparisons}
\begin{figure}[ht]
    \centering
    \includegraphics[width=10cm]{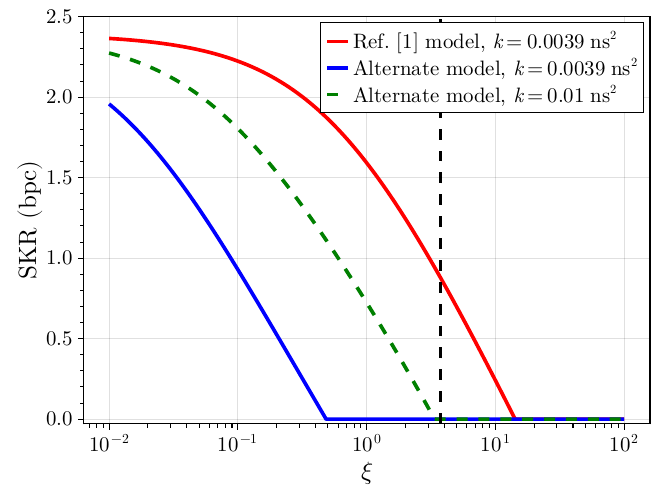}
    \caption{Plot of secure key rates (SKR) in bits per coincidence for different values of $\xi$. The parameters used are $\sigma_{coh}=1.49$ ns, $\sigma_{cor}=2.8$ ps, and $k = 0.0039\ \text{ns}^2$ based on the experiment in Ref.~\cite{PhysRevA.90.062331}. The red curve shows the predicted key rate using the Ref.~\cite{mower_2013} model and the blue curve shows the key rate using the alternate model. The dashed black line is at $\xi=3.74$, which is the value reported in the Ref.~\cite{PhysRevA.90.062331} experiment. The dashed green curve shows the key rate using the alternate model, but instead with a dispersion coefficient of $k=0.01\ \text{ns}^2$.}
    \label{fig:lee_exp}
\end{figure}

There have been multiple experiments in the past few years that have sought to facilitate secure key transfer through DO-QKD protocols, but have not been in the applicable high-dispersion regime. As an example, the parameters of the Ref.~\cite{PhysRevA.90.062331} experiment were $\sigma_{coh}=1.49$ ns, $\sigma_{cor} \sim 3$ ps, $k = 0.0039\ \text{ns}^2$ (corresponding to $|D|=1500\ \text{ps}/\text{nm}$), and a measured value of $\xi=3.74$ \cite{PhysRevA.90.062331,lee2018high}. In this experiment, the security analysis was done using the Ref.~\cite{mower_2013} model; however, we can observe that $4\sigma_{coh}\sigma_{cor} = 0.018\ \text{ns}^2 > k$. 
This means that the analysis is not attributing a sufficient amount of information to Eve. After reconciliation and subtracting off finite-key effects, the Ref.~\cite{PhysRevA.90.062331} experiment reported $I(A:B) = 2.39$ bpc and a Holevo information for Eve of $\chi_1(A:E) = 1.56$ bpc (including finite-key effects). Using the alternate model, we can show that Eve can achieve a Holevo information of $\chi_2(A:E) = 4.92$ bpc.
This implies that if we demand unconditional security, then the experiment achieves zero secure key rate. We note that in this case the Holevo information was given by $\chi(A:E)=S(\rho_E)-S(\rho_{E|T})$, as key is only generated from arrival-time measurements (with dispersed-arrival time measurements used for security). Furthermore, our alternative channel is provably non-optimal as we have found other single-mode Gaussian channels that leak even more information to Eve. These channels involve adding anti-correlated noise to Bob's submatrix of the covariance matrix while ensuring that the channel positivity condition is still met. 
In Figure \ref{fig:lee_exp}, we have plotted the secure key rates (in bpc) after subjecting the state to the Ref.~\cite{mower_2013} model and to the alternate model. 
As can be seen from the dashed black line, which corresponds to the experimentally reported value of $\xi$ in Refs.~\cite{PhysRevA.90.062331,lee2018high}, the secure key rate is zero with the alternate model as opposed to approximately 1 bpc from the model in Ref.~\cite{mower_2013}.
The green, dashed curve shows that for a larger value of dispersion, $k=0.01\ \text{ns}^2$, the key rates predicted by the alternate model approach the rates predicted by the Ref. \cite{mower_2013} model. In fact, if we take the $k\rightarrow\infty$ limit, then the key rate from the alternate model appears identical to the red curve in Fig.~\ref{fig:lee_exp}. However, according to the green curve, even after more than doubling the applied dispersion relative to the experimentally-used setting, the same correlated-variance measurements would still imply zero secure key rate.
Moreover, to recover the same key rate from the alternate model as the rate that was reported in the experiment, we would need to observe $\tilde{\xi} = 0.109$, which is significantly lower than the experimentally reported value in Refs.~\cite{PhysRevA.90.062331,lee2018high}. To get a sense for the requirements on the experimental apparatus necessary to observe a value this small, we can consider the contribution of uncertainty in the timing jitter to $\xi$. In the Ref.~\cite{PhysRevA.90.062331} experiment, the detector-timing jitter was approximately $80$ ps FWHM and an uncertainty of only $200$ fs in the timing jitter (with no other noise sources present) would lead to $\xi=3.3$. This implies that generating key by obtaining low enough values of $\xi$, while remaining in the low-dispersion limit, is very difficult with this experimental setup. Here we have used timing jitter as an example of an experimental noise source, but there are many other sources, such as photon loss and detector dark counts, that can contribute to changes in the correlated variances.

\section{Discussion of Future Prospects for DO-QKD in Finite-Dispersion Regime}
\label{sec:proposed_solutions}
The first question to consider is whether through making additional constraining measurements it is possible to rule out all valid single-mode channels other than the Ref.~\cite{mower_2013} model. To distinguish the Ref. \cite{mower_2013} model from the alternate model, we need information about $\text{Cov}(T_B,D_B)$, but there are significant obstacles to obtaining information that involves the variances of just the signal or idler photon. 
For continuous-wave pumping of the SPDC source, the mean time of photon generation ($\langle T_A \rangle$,$\langle T_B \rangle$) is unknown for a given pair and so direct measurement of $\text{Cov}(T_B,D_B)$ is not possible.
However, even in the case of a pumped-pulse laser, direct measurement of individual system variances and covariances is challenging as the $\epsilon$ corresponding to $\xi \approx 1$ (using the Ref.~\cite{mower_2013} paper values) is on the order of $10^{-4}$. This would require being able to make measurements with a timing resolution of approximately $500$ fs.

We will now consider limits to the secure key rate that could reasonably be extracted from a DO-QKD scheme. By using higher broadband SPDC sources, the correlation time can be decreased, and as long as the laser has a relatively short coherence time, then DO-QKD may work in the fashion outlined in Ref.~\cite{mower_2013}. This follows from the fact that in this limit, Alice and Bob share the TMSV state as discussed in Sec. \ref{sec_not_epr}. However, we note that the amount of dispersion that can be applied sets a strict upper bound to the number of bits that we can hope to generate in such a HD-QKD scheme. 
In general, the maximum number of bits that can be extracted from a DO-QKD scheme is approximately $\log_2\left(\sigma_{coh}/\sigma_{cor}\right)$, however, if the detector timing jitter of the instruments is greater than the correlation time, then we cannot hope to extract more than approximately $\log_2\left(\sigma_{coh}/\sigma_{J}\right)$ bits in the time-basis. In the unconditional security model, we have assumed that Eve does not have access to Alice and Bob's labs \cite{RevModPhys.84.621}. Therefore, by characterizing the timing jitter, Alice and Bob can safely subtract off the jitter's contribution to the final covariance matrix. In practice, experimentalists do this in order to show non-zero secure key rates \cite{mower_2013,PhysRevA.90.062331}.
% However, as we will shortly see, it is still required that Alice and Bob have precise knowledge of the jitter.
If we are limited to working in the physically-relevant, low-dispersion regime, we can try to ensure that there is sufficiently low uncertainty in the jitter to decrease the noise that may be attributable to Eve, and therefore the Holevo information, to a tolerable level. 
% The problem with this approach is that noise sources in the lab generally make it very challenging to constrain $\xi$ sufficiently.
% To see why this is the case, we can consider Alice and Bob's detectors having timing jitter, which prohibits exact observation of the changes in correlation time.
As an example, consider a detector with timing jitter, $\sigma_J$, and an uncertainty in the timing jitter of $\Delta_J$. The value of $\xi$, measured in the dispersed-time basis, that is a direct result of timing jitter uncertainty is 
\begin{equation*}
    \xi_J = \frac{32 \Delta_J \sigma_{coh}^2 (\Delta_J + 2 \sigma_J)}{k^2 + 16 \sigma_{coh}^2 \sigma_{cor}^2}.
\end{equation*}
If we are operating in the limit where $k\leq 4\sigma_{coh}\sigma_{cor}$, then the interference that needs to be attributed to Eve is at least $\xi_J\geq \frac{\Delta_J (\Delta_J + 2 \sigma_J)}{\sigma_{cor}^2}$. If $\sigma_J= 2\text{ ps}$, which is approximately the best experimentally achievable timing jitter for $\lambda=1550\text{ nm}$~\cite{Korzh2020}, $\sigma_{cor}= 1\text{ ps}$, and Alice and Bob have an uncertainty in their timing jitter of $\Delta_J=0.2\text{ ps}$, then we already have $\xi_J\geq 0.85$. To put this into context, if we assume a group delay slope of $|D|=10000\text{ ps/nm}$ ($k=0.026 \text{ ns}^2$), and a coherence time of $\sigma_{coh} = 6.4\ \text{ns}$, then the Holevo information is at least 1.15 bpc, which gives an upper-bound of approximately 9.85 bpc when there are no other noise sources aside from state-of-the art timing jitter.
Achieving these values would require the users to be able to determine their detector's timing jitter to an accuracy on the order of hundreds of femtoseconds, which would require more precise characterization than has been employed in experiments thus far. Calibrating the value of the timing jitter may need to be repeated for each individual device due to device-dependent fluctuations on the order of the desired uncertainty. 
An additional complication is that getting below tens of picoseconds of geometric detector timing jitter leads to significantly lower system efficiencies \cite{PhysRevApplied.19.044093, Korzh2020}.
Furthermore, for high-precision detectors, the timing jitter may not even have a Gaussian profile in practice, which creates a challenge for the security analysis of the QKD protocol.

If we instead consider an example that is more consistent with commercially available detectors, a timing jitter of $\sigma_J=10$ ps and an uncertainty of $\Delta_J=500\text{ fs}$, then we find an upper-bound on the secure key rate of 6.13 bpc, where Eve's Holevo information is $\chi(A:E)\geq 2.65$ bpc.
In practice, there are many additional noise sources that significantly decrease $I(A:B)$, such as photon loss, dark counts, and potentially the timing jitter in electronic components. This means that in practice we expect secure rates significantly lower than 6.13 bpc. To illustrate this point through an experimental example, in Ref.~\cite{PhysRevA.90.062331} a 6-bit encoding was used, but the classical key rate achieved as a result of noise, and after factoring in the reconcilliation efficiency and finite-key effects, was only $2.39$ bpc.
Furthermore, increasing the coherence time, while holding the remaining parameters fixed, achieves nearly no advantage as the Holevo information grows proportionally to the coherence time in the low-dispersion regime. Doubling the dispersion would allow for approximately 1 additional bit of secure key, and so in order to get a significantly larger photon information efficiency than $\sim 10$ bpc, one would require exponential improvements in the performance of photon detectors or the amount of dispersion that can be applied. 

\section{Conclusion}
We have demonstrated that in the finite-dispersion regime of DO-QKD, the security analysis underpinning a number of experiments and theoretical works is flawed. By proposing an alternative model, which is indistinguishable from the Ref.~\cite{mower_2013} model through the performed correlated measurements, we were able to show that the information accessible to Eve is greater than reported in multiple previous works. To guarantee unconditional security in the experimentally-relevant dispersion regime, it becomes necessary to perform a maximization of Eve's Holevo information over all covariance matrices consistent with experimental outcomes. 
The maximized Holevo information is at least as large as the Holevo information calculated using our alternative model, and we have found cases where the optimized quantity in fact gives Eve more information.
While DO-QKD remains a theoretically viable means of distributing secure cryptographic keys, future experiments likely must either use detectors with significantly lower timing jitter or use dispersive elements that apply greater amounts of dispersion than have been used to date.
An open question is whether there is a simple, analytic form that describes the best collective attack that Eve can implement, subject to the constraints associated with Alice's and Bob's security protocol.
The field is still missing a definitive experimental verification that secure key can be generated through the DO-QKD protocol, when Alice's transmission to Bob is subjected to arbitrary collective attacks. As higher values of dispersion are challenging to obtain, an experiment which uses a short coherence time, short correlation time, and state-of-the-art detectors may be able to achieve secure key transfer. 
As a final note, the authors urge further caution when analyzing QKD schemes for unconditional security since even after taking into account uncertainty in timing jitter, we have not addressed noise sources such as photon loss and dark counts, which may mask Eve's interference if not handled properly.

\section{Acknowledgements}
The authors would like to thank Akshay Seshadri, Michael Grayson, Dileep Reddy, Emanuel Knill, Scott Glancy, Murat Can Sarihan, and Chee Wei Wong for helpful suggestions and discussions.
This material is based upon work supported by the Army Research Office Grant No: W911NF-21-2-0214 (ARO-MURI).
At the time this work was performed, A. Kwiatkowski was supported by the National Institute of Standards and Technology, and A. Kyle was supported by the by the Army Research Office grant W911NF-23-1-0376 and National Science Foundation Quantum Leap Challenge Institutes (QLCI) award OMA–2016244.

\appendix
\section*{Appendix}
\section{Physical Interpretation of Ref.~\cite{mower_2013} Model}
\label{app:physical_mower}
A phase-insensitive thermal, lossy Gaussian channel can be parameterized by its transmissivity, $\eta$, and by an excess noise factor, $\epsilon$ \cite{RevModPhys.84.621,PhysRevResearch.2.013208}. Eve can act with arbitrary local Gaussian unitaries on Bob's system before and after the application of the channel. A reasonable channel to consider is the one where Eve applies the diagonalizing symplectic transform $S$ to Bob's system, followed by the thermal lossy phase-insensitive channel $\mathcal{N}(\eta,\epsilon):\mathbf{T_0}=\sqrt{\eta}\ \mathbf{I}, \mathbf{N_0}=\frac{1}{2}(1-\eta+\epsilon)\mathbf{I}$ with $\eta\in(0,1)$, $\epsilon\in(0,\infty)$, and then applies the inverse symplectic operation ($S^{-1}$). This attack is motivated by the fact that in the case where Alice and Bob start with a TMSV state, then the diagonalizing symplectic is trivial and Eve just applies the optimal attack on the TMSV state. 
After this sequence of operations, the state takes the form
\begin{equation*}
    \Gamma' = \begin{pmatrix}
        \gamma_{AA} & \sqrt{\eta}\gamma_{AB} \\
        \sqrt{\eta}\gamma_{BA} & \left(\eta +\frac{4(1 + \epsilon - \eta)}{
 4 \sigma_{coh}^2 + \sigma_{cor}^2}\sigma_{coh} \sigma_{cor}\right)\gamma_{BB} \\
    \end{pmatrix}.
\end{equation*}
We can define $\eta'\equiv (1-\sqrt{\eta})$ and $\epsilon'\equiv \frac{4 (1 + \epsilon - \eta) \sigma_{coh} \sigma_{cor}}{
4 \sigma_{coh}^2 + \sigma_{cor}^2}-(1-\eta)$ so that the prefactors in front of $\gamma_{AB}$ and $\gamma_{BB}$ becomes $(1-\eta')$ and $(1+\epsilon')$, respectively. The parameter ranges for the new coefficients are $\eta'\in(0,1)$ and $\epsilon'\in(-1 + \frac{4 \sigma_{coh} \sigma_{cor}}{4 \sigma_{coh}^2 + \sigma_{cor}^2},\infty)$. If we instead had applied a thermal amplifier channel $\mathcal{N}(\eta,\epsilon):\mathbf{T_0}=\sqrt{\eta},\ \mathbf{N_0}=\frac{1}{2}(\eta-1+\epsilon)\mathbf{I}$ with $\eta\in(1,\infty)$, $\epsilon\in(0,\infty)$, then we still have that the covariance matrix transforms as under the Ref.~\cite{mower_2013} model. In this case, the relevant parameter regime is $\eta'\in(-\infty,0)$ and $\epsilon'\in(0,\infty)$. Therefore, we have demonstrated that if the values of the model in Ref.~\cite{mower_2013} lie in the range from $\eta'\in(-\infty,1)$ and $\epsilon'\in(-1 + \frac{4 \sigma_{coh} \sigma_{cor}}{4 \sigma_{coh}^2 + \sigma_{cor}^2},\infty)$, then the model corresponds to applying a phase-insensitive channel in the basis where the state is a TMSV state. 
% \begin{equation*}
%     \Gamma_{d}'=(\mathbf{I}\oplus\mathbf{T_0})\Gamma(\mathbf{I}\oplus\mathbf{T_0^T})+\mathbf{0}\oplus\mathbf{N_0}=\begin{pmatrix}
%         \gamma_{AA} & (1-\eta)\gamma_{AB} \\
%         (1-\eta)\gamma_{BA} & (1-\eta)^2\gamma_{BB} + \mathbf{N_0}
%         \end{pmatrix}
% \end{equation*}
\section{Changes of Correlated Variances in Different Bases}
\label{app:correlated_variances}
The following are the list of all changes in correlated variances as a result of Eve's attack, depending on which basis both Alice and Bob measure. For completeness, we list the variance changes of all sums and differences of the correlated measurement outcomes that Alice and Bob make, which are equivalent for both the model of Ref.~\cite{mower_2013} and our alternative collective attack. Here we will we use the notation $\Delta'(x)\equiv \text{Var}(x')/\text{Var}(x)$, where $x$ is an arbitrary linear combination of the $T$ and $D$ observables.
\begin{gather*}
\begin{aligned}
\Delta'(T_A-T_B) &= 1 + \xi \\
\Delta'\left(\frac{k}{2}D_A+\frac{k}{2}D_B\right) &= 1 + \xi \\
\Delta'\left(T_A+\frac{k}{2}D_B\right) &= \frac{64 \sigma_{coh}^2 \sigma_{cor}^4(1+\xi)+k^2((4-8 \eta)\sigma_{coh}^2 + (1+2 \eta + 4 \xi) \sigma_{cor}^2)}{64 \sigma_{coh}^2 \sigma_{cor}^4 + k^2 (4 \sigma_{coh}^2 + \sigma_{cor}^2)} \\
\Delta'\left(T_B-\frac{k}{2}D_A\right) &= \frac{64 (1+\xi) \sigma_{coh}^2 \sigma_{cor}^4+k^2 (4 \sigma_{coh}^2 + \sigma_{cor}^2)}{64\sigma_{coh}^2 \sigma_{cor}^4 + k^2 (4 \sigma_{coh}^2 + \sigma_{cor}^2)} \\
\Delta'(T_A+T_B) &= 1 - \eta + \frac{(\eta + \xi) \sigma_{cor}^2}{4 \sigma_{coh}^2} \\
\Delta'\left(\frac{k}{2}D_A-\frac{k}{2}D_B\right) &= 1 - \eta + \frac{(\eta + \xi) \sigma_{cor}^2}{4 \sigma_{coh}^2} \\
\end{aligned}
\end{gather*}
\begin{gather*}
\begin{aligned}
\Delta'\left(T_A-\frac{k}{2}D_B\right) &= 
\frac{64 \sigma_{coh}^2 \sigma_{cor}^2 \left(\sigma_{cor}^2 (\eta +\xi)-4(\eta-1) \sigma_{coh}^2\right)+k^2
\left(\sigma_{cor}^2 (1+2 \eta +4 \xi)+(4-8 \eta ) \sigma_{coh}^2\right)}{k^2 \left(4\sigma_{coh}^2+\sigma_{cor}^2\right)+256\sigma_{coh}^4 \sigma_{cor}^2} \\
\Delta'\left(T_B+\frac{k}{2}D_A\right) &= \frac{64 \sigma_{coh}^2 \sigma_{cor}^2 \left(\sigma_{cor}^2 (\eta +\xi)-4 (\eta-1)\sigma_{coh}^2\right)+k^2\left(4 \sigma_{coh}^2+\sigma_{cor}^2\right)}{k^2 \left(4 \sigma_{coh}^2+\sigma_{cor}^2\right)+256
\sigma_{coh}^4\sigma_{cor}^2}
\end{aligned}
\end{gather*}
The $\pm$ signs in the above equations are a result of Alice and Bob applying normal and anomalous dispersion, respectively.

\bibliography{ref.bib}

\begin{thebibliography}{44}%
\makeatletter
\providecommand \@ifxundefined [1]{%
 \@ifx{#1\undefined}
}%
\providecommand \@ifnum [1]{%
 \ifnum #1\expandafter \@firstoftwo
 \else \expandafter \@secondoftwo
 \fi
}%
\providecommand \@ifx [1]{%
 \ifx #1\expandafter \@firstoftwo
 \else \expandafter \@secondoftwo
 \fi
}%
\providecommand \natexlab [1]{#1}%
\providecommand \enquote  [1]{``#1''}%
\providecommand \bibnamefont  [1]{#1}%
\providecommand \bibfnamefont [1]{#1}%
\providecommand \citenamefont [1]{#1}%
\providecommand \href@noop [0]{\@secondoftwo}%
\providecommand \href [0]{\begingroup \@sanitize@url \@href}%
\providecommand \@href[1]{\@@startlink{#1}\@@href}%
\providecommand \@@href[1]{\endgroup#1\@@endlink}%
\providecommand \@sanitize@url [0]{\catcode `\\12\catcode `\$12\catcode `\&12\catcode `\#12\catcode `\^12\catcode `\_12\catcode `\%12\relax}%
\providecommand \@@startlink[1]{}%
\providecommand \@@endlink[0]{}%
\providecommand \url  [0]{\begingroup\@sanitize@url \@url }%
\providecommand \@url [1]{\endgroup\@href {#1}{\urlprefix }}%
\providecommand \urlprefix  [0]{URL }%
\providecommand \Eprint [0]{\href }%
\providecommand \doibase [0]{https://doi.org/}%
\providecommand \selectlanguage [0]{\@gobble}%
\providecommand \bibinfo  [0]{\@secondoftwo}%
\providecommand \bibfield  [0]{\@secondoftwo}%
\providecommand \translation [1]{[#1]}%
\providecommand \BibitemOpen [0]{}%
\providecommand \bibitemStop [0]{}%
\providecommand \bibitemNoStop [0]{.\EOS\space}%
\providecommand \EOS [0]{\spacefactor3000\relax}%
\providecommand \BibitemShut  [1]{\csname bibitem#1\endcsname}%
\let\auto@bib@innerbib\@empty
%</preamble>
\bibitem [{\citenamefont {Mower}\ \emph {et~al.}(2013)\citenamefont {Mower}, \citenamefont {Zhang}, \citenamefont {Desjardins}, \citenamefont {Lee}, \citenamefont {Shapiro},\ and\ \citenamefont {Englund}}]{mower_2013}%
  \BibitemOpen
  \bibfield  {author} {\bibinfo {author} {\bibfnamefont {J.}~\bibnamefont {Mower}}, \bibinfo {author} {\bibfnamefont {Z.}~\bibnamefont {Zhang}}, \bibinfo {author} {\bibfnamefont {P.}~\bibnamefont {Desjardins}}, \bibinfo {author} {\bibfnamefont {C.}~\bibnamefont {Lee}}, \bibinfo {author} {\bibfnamefont {J.~H.}\ \bibnamefont {Shapiro}},\ and\ \bibinfo {author} {\bibfnamefont {D.}~\bibnamefont {Englund}},\ }\bibfield  {title} {\bibinfo {title} {High-dimensional quantum key distribution using dispersive optics},\ }\href {https://doi.org/10.1103/PhysRevA.87.062322} {\bibfield  {journal} {\bibinfo  {journal} {Phys. Rev. A}\ }\textbf {\bibinfo {volume} {87}},\ \bibinfo {pages} {062322} (\bibinfo {year} {2013})}\BibitemShut {NoStop}%
\bibitem [{\citenamefont {Lee}\ \emph {et~al.}(2014)\citenamefont {Lee}, \citenamefont {Zhang}, \citenamefont {Steinbrecher}, \citenamefont {Zhou}, \citenamefont {Mower}, \citenamefont {Zhong}, \citenamefont {Wang}, \citenamefont {Hu}, \citenamefont {Horansky}, \citenamefont {Verma}, \citenamefont {Lita}, \citenamefont {Mirin}, \citenamefont {Marsili}, \citenamefont {Shaw}, \citenamefont {Nam}, \citenamefont {Wornell}, \citenamefont {Wong}, \citenamefont {Shapiro},\ and\ \citenamefont {Englund}}]{PhysRevA.90.062331}%
  \BibitemOpen
  \bibfield  {author} {\bibinfo {author} {\bibfnamefont {C.}~\bibnamefont {Lee}}, \bibinfo {author} {\bibfnamefont {Z.}~\bibnamefont {Zhang}}, \bibinfo {author} {\bibfnamefont {G.~R.}\ \bibnamefont {Steinbrecher}}, \bibinfo {author} {\bibfnamefont {H.}~\bibnamefont {Zhou}}, \bibinfo {author} {\bibfnamefont {J.}~\bibnamefont {Mower}}, \bibinfo {author} {\bibfnamefont {T.}~\bibnamefont {Zhong}}, \bibinfo {author} {\bibfnamefont {L.}~\bibnamefont {Wang}}, \bibinfo {author} {\bibfnamefont {X.}~\bibnamefont {Hu}}, \bibinfo {author} {\bibfnamefont {R.~D.}\ \bibnamefont {Horansky}}, \bibinfo {author} {\bibfnamefont {V.~B.}\ \bibnamefont {Verma}}, \bibinfo {author} {\bibfnamefont {A.~E.}\ \bibnamefont {Lita}}, \bibinfo {author} {\bibfnamefont {R.~P.}\ \bibnamefont {Mirin}}, \bibinfo {author} {\bibfnamefont {F.}~\bibnamefont {Marsili}}, \bibinfo {author} {\bibfnamefont {M.~D.}\ \bibnamefont {Shaw}}, \bibinfo {author} {\bibfnamefont {S.~W.}\ \bibnamefont {Nam}}, \bibinfo {author} {\bibfnamefont {G.~W.}\ \bibnamefont
  {Wornell}}, \bibinfo {author} {\bibfnamefont {F.~N.~C.}\ \bibnamefont {Wong}}, \bibinfo {author} {\bibfnamefont {J.~H.}\ \bibnamefont {Shapiro}},\ and\ \bibinfo {author} {\bibfnamefont {D.}~\bibnamefont {Englund}},\ }\bibfield  {title} {\bibinfo {title} {Entanglement-based quantum communication secured by nonlocal dispersion cancellation},\ }\href {https://doi.org/10.1103/PhysRevA.90.062331} {\bibfield  {journal} {\bibinfo  {journal} {Phys. Rev. A}\ }\textbf {\bibinfo {volume} {90}},\ \bibinfo {pages} {062331} (\bibinfo {year} {2014})}\BibitemShut {NoStop}%
\bibitem [{\citenamefont {Liu}\ \emph {et~al.}(2019)\citenamefont {Liu}, \citenamefont {Yao}, \citenamefont {Wang}, \citenamefont {Li}, \citenamefont {Wang}, \citenamefont {You}, \citenamefont {Huang},\ and\ \citenamefont {Zhang}}]{10.1063/1.5089784}%
  \BibitemOpen
  \bibfield  {author} {\bibinfo {author} {\bibfnamefont {X.}~\bibnamefont {Liu}}, \bibinfo {author} {\bibfnamefont {X.}~\bibnamefont {Yao}}, \bibinfo {author} {\bibfnamefont {H.}~\bibnamefont {Wang}}, \bibinfo {author} {\bibfnamefont {H.}~\bibnamefont {Li}}, \bibinfo {author} {\bibfnamefont {Z.}~\bibnamefont {Wang}}, \bibinfo {author} {\bibfnamefont {L.}~\bibnamefont {You}}, \bibinfo {author} {\bibfnamefont {Y.}~\bibnamefont {Huang}},\ and\ \bibinfo {author} {\bibfnamefont {W.}~\bibnamefont {Zhang}},\ }\bibfield  {title} {\bibinfo {title} {{Energy-time entanglement-based dispersive optics quantum key distribution over optical fibers of 20 km}},\ }\href {https://doi.org/10.1063/1.5089784} {\bibfield  {journal} {\bibinfo  {journal} {Applied Physics Letters}\ }\textbf {\bibinfo {volume} {114}},\ \bibinfo {pages} {141104} (\bibinfo {year} {2019})}\BibitemShut {NoStop}%
\bibitem [{\citenamefont {Liu}\ \emph {et~al.}(2020)\citenamefont {Liu}, \citenamefont {Yao}, \citenamefont {Xue}, \citenamefont {Wang}, \citenamefont {Li}, \citenamefont {Wang}, \citenamefont {You}, \citenamefont {Feng}, \citenamefont {Liu}, \citenamefont {Cui}, \citenamefont {Huang},\ and\ \citenamefont {Zhang}}]{10.1063/5.0002595}%
  \BibitemOpen
  \bibfield  {author} {\bibinfo {author} {\bibfnamefont {X.}~\bibnamefont {Liu}}, \bibinfo {author} {\bibfnamefont {X.}~\bibnamefont {Yao}}, \bibinfo {author} {\bibfnamefont {R.}~\bibnamefont {Xue}}, \bibinfo {author} {\bibfnamefont {H.}~\bibnamefont {Wang}}, \bibinfo {author} {\bibfnamefont {H.}~\bibnamefont {Li}}, \bibinfo {author} {\bibfnamefont {Z.}~\bibnamefont {Wang}}, \bibinfo {author} {\bibfnamefont {L.}~\bibnamefont {You}}, \bibinfo {author} {\bibfnamefont {X.}~\bibnamefont {Feng}}, \bibinfo {author} {\bibfnamefont {F.}~\bibnamefont {Liu}}, \bibinfo {author} {\bibfnamefont {K.}~\bibnamefont {Cui}}, \bibinfo {author} {\bibfnamefont {Y.}~\bibnamefont {Huang}},\ and\ \bibinfo {author} {\bibfnamefont {W.}~\bibnamefont {Zhang}},\ }\bibfield  {title} {\bibinfo {title} {{An entanglement-based quantum network based on symmetric dispersive optics quantum key distribution}},\ }\href {https://doi.org/10.1063/5.0002595} {\bibfield  {journal} {\bibinfo  {journal} {APL Photonics}\ }\textbf {\bibinfo {volume}
  {5}},\ \bibinfo {pages} {076104} (\bibinfo {year} {2020})}\BibitemShut {NoStop}%
\bibitem [{\citenamefont {Liu}\ \emph {et~al.}(2023)\citenamefont {Liu}, \citenamefont {Lin}, \citenamefont {Liu}, \citenamefont {Feng}, \citenamefont {Liu}, \citenamefont {Cui}, \citenamefont {Huang},\ and\ \citenamefont {Zhang}}]{10.1088/2058-9565/acfe37}%
  \BibitemOpen
  \bibfield  {author} {\bibinfo {author} {\bibfnamefont {J.}~\bibnamefont {Liu}}, \bibinfo {author} {\bibfnamefont {Z.}~\bibnamefont {Lin}}, \bibinfo {author} {\bibfnamefont {D.}~\bibnamefont {Liu}}, \bibinfo {author} {\bibfnamefont {X.}~\bibnamefont {Feng}}, \bibinfo {author} {\bibfnamefont {F.}~\bibnamefont {Liu}}, \bibinfo {author} {\bibfnamefont {K.}~\bibnamefont {Cui}}, \bibinfo {author} {\bibfnamefont {Y.}~\bibnamefont {Huang}},\ and\ \bibinfo {author} {\bibfnamefont {W.}~\bibnamefont {Zhang}},\ }\bibfield  {title} {\bibinfo {title} {{High-dimensional quantum key distribution using energy-time entanglement over 242 km partially deployed fiber}},\ }\href {https://doi.org/10.1088/2058-9565/acfe37} {\bibfield  {journal} {\bibinfo  {journal} {Quantum Science and Technology}\ }\textbf {\bibinfo {volume} {9}},\ \bibinfo {pages} {015003} (\bibinfo {year} {2023})}\BibitemShut {NoStop}%
\bibitem [{\citenamefont {Chang}\ \emph {et~al.}(2023)\citenamefont {Chang}, \citenamefont {Sarihan}, \citenamefont {Cheng}, \citenamefont {Zhang},\ and\ \citenamefont {Wong}}]{10.1088/2058-9565/ad0f6f}%
  \BibitemOpen
  \bibfield  {author} {\bibinfo {author} {\bibfnamefont {K.-C.}\ \bibnamefont {Chang}}, \bibinfo {author} {\bibfnamefont {M.~C.}\ \bibnamefont {Sarihan}}, \bibinfo {author} {\bibfnamefont {X.}~\bibnamefont {Cheng}}, \bibinfo {author} {\bibfnamefont {Z.}~\bibnamefont {Zhang}},\ and\ \bibinfo {author} {\bibfnamefont {C.~W.}\ \bibnamefont {Wong}},\ }\bibfield  {title} {\bibinfo {title} {{Large-alphabet time-bin quantum key distribution and Einstein–Podolsky–Rosen steering via dispersive optics}},\ }\href {https://doi.org/10.1088/2058-9565/ad0f6f} {\bibfield  {journal} {\bibinfo  {journal} {Quantum Science and Technology}\ }\textbf {\bibinfo {volume} {9}},\ \bibinfo {pages} {015018} (\bibinfo {year} {2023})}\BibitemShut {NoStop}%
\bibitem [{\citenamefont {Bennett}\ and\ \citenamefont {Brassard}(2014)}]{BENNETT20147}%
  \BibitemOpen
  \bibfield  {author} {\bibinfo {author} {\bibfnamefont {C.~H.}\ \bibnamefont {Bennett}}\ and\ \bibinfo {author} {\bibfnamefont {G.}~\bibnamefont {Brassard}},\ }\bibfield  {title} {\bibinfo {title} {Quantum cryptography: Public key distribution and coin tossing},\ }\href {https://doi.org/https://doi.org/10.1016/j.tcs.2014.05.025} {\bibfield  {journal} {\bibinfo  {journal} {Theoretical Computer Science}\ }\textbf {\bibinfo {volume} {560}},\ \bibinfo {pages} {7} (\bibinfo {year} {2014})},\ \bibinfo {note} {theoretical Aspects of Quantum Cryptography – celebrating 30 years of BB84}\BibitemShut {NoStop}%
\bibitem [{\citenamefont {Ekert}(1991)}]{PhysRevLett.67.661}%
  \BibitemOpen
  \bibfield  {author} {\bibinfo {author} {\bibfnamefont {A.~K.}\ \bibnamefont {Ekert}},\ }\bibfield  {title} {\bibinfo {title} {Quantum cryptography based on bell's theorem},\ }\href {https://doi.org/10.1103/PhysRevLett.67.661} {\bibfield  {journal} {\bibinfo  {journal} {Phys. Rev. Lett.}\ }\textbf {\bibinfo {volume} {67}},\ \bibinfo {pages} {661} (\bibinfo {year} {1991})}\BibitemShut {NoStop}%
\bibitem [{\citenamefont {Etcheverry}\ \emph {et~al.}(2013)\citenamefont {Etcheverry}, \citenamefont {Ca{\~{n}}as}, \citenamefont {G{\'o}mez}, \citenamefont {Nogueira}, \citenamefont {Saavedra}, \citenamefont {Xavier},\ and\ \citenamefont {Lima}}]{Etcheverry2013}%
  \BibitemOpen
  \bibfield  {author} {\bibinfo {author} {\bibfnamefont {S.}~\bibnamefont {Etcheverry}}, \bibinfo {author} {\bibfnamefont {G.}~\bibnamefont {Ca{\~{n}}as}}, \bibinfo {author} {\bibfnamefont {E.~S.}\ \bibnamefont {G{\'o}mez}}, \bibinfo {author} {\bibfnamefont {W.~A.~T.}\ \bibnamefont {Nogueira}}, \bibinfo {author} {\bibfnamefont {C.}~\bibnamefont {Saavedra}}, \bibinfo {author} {\bibfnamefont {G.~B.}\ \bibnamefont {Xavier}},\ and\ \bibinfo {author} {\bibfnamefont {G.}~\bibnamefont {Lima}},\ }\bibfield  {title} {\bibinfo {title} {Quantum key distribution session with 16-dimensional photonic states},\ }\href {https://doi.org/10.1038/srep02316} {\bibfield  {journal} {\bibinfo  {journal} {Scientific Reports}\ }\textbf {\bibinfo {volume} {3}},\ \bibinfo {pages} {2316} (\bibinfo {year} {2013})}\BibitemShut {NoStop}%
\bibitem [{\citenamefont {Walborn}\ \emph {et~al.}(2006)\citenamefont {Walborn}, \citenamefont {Lemelle}, \citenamefont {Almeida},\ and\ \citenamefont {Ribeiro}}]{PhysRevLett.96.090501}%
  \BibitemOpen
  \bibfield  {author} {\bibinfo {author} {\bibfnamefont {S.~P.}\ \bibnamefont {Walborn}}, \bibinfo {author} {\bibfnamefont {D.~S.}\ \bibnamefont {Lemelle}}, \bibinfo {author} {\bibfnamefont {M.~P.}\ \bibnamefont {Almeida}},\ and\ \bibinfo {author} {\bibfnamefont {P.~H.~S.}\ \bibnamefont {Ribeiro}},\ }\bibfield  {title} {\bibinfo {title} {Quantum key distribution with higher-order alphabets using spatially encoded qudits},\ }\href {https://doi.org/10.1103/PhysRevLett.96.090501} {\bibfield  {journal} {\bibinfo  {journal} {Phys. Rev. Lett.}\ }\textbf {\bibinfo {volume} {96}},\ \bibinfo {pages} {090501} (\bibinfo {year} {2006})}\BibitemShut {NoStop}%
\bibitem [{\citenamefont {Zhang}\ \emph {et~al.}(2014)\citenamefont {Zhang}, \citenamefont {Mower}, \citenamefont {Englund}, \citenamefont {Wong},\ and\ \citenamefont {Shapiro}}]{PhysRevLett.112.120506}%
  \BibitemOpen
  \bibfield  {author} {\bibinfo {author} {\bibfnamefont {Z.}~\bibnamefont {Zhang}}, \bibinfo {author} {\bibfnamefont {J.}~\bibnamefont {Mower}}, \bibinfo {author} {\bibfnamefont {D.}~\bibnamefont {Englund}}, \bibinfo {author} {\bibfnamefont {F.~N.~C.}\ \bibnamefont {Wong}},\ and\ \bibinfo {author} {\bibfnamefont {J.~H.}\ \bibnamefont {Shapiro}},\ }\bibfield  {title} {\bibinfo {title} {Unconditional security of time-energy entanglement quantum key distribution using dual-basis interferometry},\ }\href {https://doi.org/10.1103/PhysRevLett.112.120506} {\bibfield  {journal} {\bibinfo  {journal} {Phys. Rev. Lett.}\ }\textbf {\bibinfo {volume} {112}},\ \bibinfo {pages} {120506} (\bibinfo {year} {2014})}\BibitemShut {NoStop}%
\bibitem [{\citenamefont {Nunn}\ \emph {et~al.}(2013)\citenamefont {Nunn}, \citenamefont {Wright}, \citenamefont {S\"{o}ller}, \citenamefont {Zhang}, \citenamefont {Walmsley},\ and\ \citenamefont {Smith}}]{Nunn:13}%
  \BibitemOpen
  \bibfield  {author} {\bibinfo {author} {\bibfnamefont {J.}~\bibnamefont {Nunn}}, \bibinfo {author} {\bibfnamefont {L.~J.}\ \bibnamefont {Wright}}, \bibinfo {author} {\bibfnamefont {C.}~\bibnamefont {S\"{o}ller}}, \bibinfo {author} {\bibfnamefont {L.}~\bibnamefont {Zhang}}, \bibinfo {author} {\bibfnamefont {I.~A.}\ \bibnamefont {Walmsley}},\ and\ \bibinfo {author} {\bibfnamefont {B.~J.}\ \bibnamefont {Smith}},\ }\bibfield  {title} {\bibinfo {title} {Large-alphabet time-frequency entangled quantum key distribution by means of time-to-frequency conversion},\ }\href {https://doi.org/10.1364/OE.21.015959} {\bibfield  {journal} {\bibinfo  {journal} {Opt. Express}\ }\textbf {\bibinfo {volume} {21}},\ \bibinfo {pages} {15959} (\bibinfo {year} {2013})}\BibitemShut {NoStop}%
\bibitem [{\citenamefont {Grosshans}\ and\ \citenamefont {Grangier}(2002)}]{PhysRevLett.88.057902}%
  \BibitemOpen
  \bibfield  {author} {\bibinfo {author} {\bibfnamefont {F.}~\bibnamefont {Grosshans}}\ and\ \bibinfo {author} {\bibfnamefont {P.}~\bibnamefont {Grangier}},\ }\bibfield  {title} {\bibinfo {title} {Continuous variable quantum cryptography using coherent states},\ }\href {https://doi.org/10.1103/PhysRevLett.88.057902} {\bibfield  {journal} {\bibinfo  {journal} {Phys. Rev. Lett.}\ }\textbf {\bibinfo {volume} {88}},\ \bibinfo {pages} {057902} (\bibinfo {year} {2002})}\BibitemShut {NoStop}%
\bibitem [{\citenamefont {Cerf}\ \emph {et~al.}(2001)\citenamefont {Cerf}, \citenamefont {L\'evy},\ and\ \citenamefont {Assche}}]{PhysRevA.63.052311}%
  \BibitemOpen
  \bibfield  {author} {\bibinfo {author} {\bibfnamefont {N.~J.}\ \bibnamefont {Cerf}}, \bibinfo {author} {\bibfnamefont {M.}~\bibnamefont {L\'evy}},\ and\ \bibinfo {author} {\bibfnamefont {G.~V.}\ \bibnamefont {Assche}},\ }\bibfield  {title} {\bibinfo {title} {Quantum distribution of gaussian keys using squeezed states},\ }\href {https://doi.org/10.1103/PhysRevA.63.052311} {\bibfield  {journal} {\bibinfo  {journal} {Phys. Rev. A}\ }\textbf {\bibinfo {volume} {63}},\ \bibinfo {pages} {052311} (\bibinfo {year} {2001})}\BibitemShut {NoStop}%
\bibitem [{\citenamefont {Weedbrook}\ \emph {et~al.}(2004)\citenamefont {Weedbrook}, \citenamefont {Lance}, \citenamefont {Bowen}, \citenamefont {Symul}, \citenamefont {Ralph},\ and\ \citenamefont {Lam}}]{PhysRevLett.93.170504}%
  \BibitemOpen
  \bibfield  {author} {\bibinfo {author} {\bibfnamefont {C.}~\bibnamefont {Weedbrook}}, \bibinfo {author} {\bibfnamefont {A.~M.}\ \bibnamefont {Lance}}, \bibinfo {author} {\bibfnamefont {W.~P.}\ \bibnamefont {Bowen}}, \bibinfo {author} {\bibfnamefont {T.}~\bibnamefont {Symul}}, \bibinfo {author} {\bibfnamefont {T.~C.}\ \bibnamefont {Ralph}},\ and\ \bibinfo {author} {\bibfnamefont {P.~K.}\ \bibnamefont {Lam}},\ }\bibfield  {title} {\bibinfo {title} {Quantum cryptography without switching},\ }\href {https://doi.org/10.1103/PhysRevLett.93.170504} {\bibfield  {journal} {\bibinfo  {journal} {Phys. Rev. Lett.}\ }\textbf {\bibinfo {volume} {93}},\ \bibinfo {pages} {170504} (\bibinfo {year} {2004})}\BibitemShut {NoStop}%
\bibitem [{\citenamefont {Grosshans}\ \emph {et~al.}(2003{\natexlab{a}})\citenamefont {Grosshans}, \citenamefont {Van~Assche}, \citenamefont {Wenger}, \citenamefont {Brouri}, \citenamefont {Cerf},\ and\ \citenamefont {Grangier}}]{Grosshans2003}%
  \BibitemOpen
  \bibfield  {author} {\bibinfo {author} {\bibfnamefont {F.}~\bibnamefont {Grosshans}}, \bibinfo {author} {\bibfnamefont {G.}~\bibnamefont {Van~Assche}}, \bibinfo {author} {\bibfnamefont {J.}~\bibnamefont {Wenger}}, \bibinfo {author} {\bibfnamefont {R.}~\bibnamefont {Brouri}}, \bibinfo {author} {\bibfnamefont {N.~J.}\ \bibnamefont {Cerf}},\ and\ \bibinfo {author} {\bibfnamefont {P.}~\bibnamefont {Grangier}},\ }\bibfield  {title} {\bibinfo {title} {Quantum key distribution using gaussian-modulated coherent states},\ }\href {https://doi.org/10.1038/nature01289} {\bibfield  {journal} {\bibinfo  {journal} {Nature}\ }\textbf {\bibinfo {volume} {421}},\ \bibinfo {pages} {238} (\bibinfo {year} {2003}{\natexlab{a}})}\BibitemShut {NoStop}%
\bibitem [{\citenamefont {Zhang}\ \emph {et~al.}(2008)\citenamefont {Zhang}, \citenamefont {Silberhorn},\ and\ \citenamefont {Walmsley}}]{PhysRevLett.100.110504}%
  \BibitemOpen
  \bibfield  {author} {\bibinfo {author} {\bibfnamefont {L.}~\bibnamefont {Zhang}}, \bibinfo {author} {\bibfnamefont {C.}~\bibnamefont {Silberhorn}},\ and\ \bibinfo {author} {\bibfnamefont {I.~A.}\ \bibnamefont {Walmsley}},\ }\bibfield  {title} {\bibinfo {title} {Secure quantum key distribution using continuous variables of single photons},\ }\href {https://doi.org/10.1103/PhysRevLett.100.110504} {\bibfield  {journal} {\bibinfo  {journal} {Phys. Rev. Lett.}\ }\textbf {\bibinfo {volume} {100}},\ \bibinfo {pages} {110504} (\bibinfo {year} {2008})}\BibitemShut {NoStop}%
\bibitem [{\citenamefont {Qi}(2006)}]{Qi:06}%
  \BibitemOpen
  \bibfield  {author} {\bibinfo {author} {\bibfnamefont {B.}~\bibnamefont {Qi}},\ }\bibfield  {title} {\bibinfo {title} {Single-photon continuous-variable quantum key distribution based on the energy-time uncertainty relation},\ }\href {https://doi.org/10.1364/OL.31.002795} {\bibfield  {journal} {\bibinfo  {journal} {Opt. Lett.}\ }\textbf {\bibinfo {volume} {31}},\ \bibinfo {pages} {2795} (\bibinfo {year} {2006})}\BibitemShut {NoStop}%
\bibitem [{\citenamefont {Fabre}\ \emph {et~al.}(2022)\citenamefont {Fabre}, \citenamefont {Keller},\ and\ \citenamefont {Milman}}]{PhysRevA.105.052429}%
  \BibitemOpen
  \bibfield  {author} {\bibinfo {author} {\bibfnamefont {N.}~\bibnamefont {Fabre}}, \bibinfo {author} {\bibfnamefont {A.}~\bibnamefont {Keller}},\ and\ \bibinfo {author} {\bibfnamefont {P.}~\bibnamefont {Milman}},\ }\bibfield  {title} {\bibinfo {title} {Time and frequency as quantum continuous variables},\ }\href {https://doi.org/10.1103/PhysRevA.105.052429} {\bibfield  {journal} {\bibinfo  {journal} {Phys. Rev. A}\ }\textbf {\bibinfo {volume} {105}},\ \bibinfo {pages} {052429} (\bibinfo {year} {2022})}\BibitemShut {NoStop}%
\bibitem [{\citenamefont {Garcia-Patron~Sanchez}(2007)}]{raul_thesis}%
  \BibitemOpen
  \bibfield  {author} {\bibinfo {author} {\bibfnamefont {R.}~\bibnamefont {Garcia-Patron~Sanchez}},\ }\emph {\bibinfo {title} {Quantum information with optical continuous variables: from Bell tests to key distribution}},\ \href@noop {} {Ph.D. thesis},\ \bibinfo  {school} {Universit\'e Libre de Bruxelles} (\bibinfo {year} {2007})\BibitemShut {NoStop}%
\bibitem [{\citenamefont {Bennett}\ \emph {et~al.}(1992)\citenamefont {Bennett}, \citenamefont {Brassard},\ and\ \citenamefont {Mermin}}]{PhysRevLett.68.557}%
  \BibitemOpen
  \bibfield  {author} {\bibinfo {author} {\bibfnamefont {C.~H.}\ \bibnamefont {Bennett}}, \bibinfo {author} {\bibfnamefont {G.}~\bibnamefont {Brassard}},\ and\ \bibinfo {author} {\bibfnamefont {N.~D.}\ \bibnamefont {Mermin}},\ }\bibfield  {title} {\bibinfo {title} {Quantum cryptography without bell's theorem},\ }\href {https://doi.org/10.1103/PhysRevLett.68.557} {\bibfield  {journal} {\bibinfo  {journal} {Phys. Rev. Lett.}\ }\textbf {\bibinfo {volume} {68}},\ \bibinfo {pages} {557} (\bibinfo {year} {1992})}\BibitemShut {NoStop}%
\bibitem [{\citenamefont {Grosshans}\ \emph {et~al.}(2003{\natexlab{b}})\citenamefont {Grosshans}, \citenamefont {Cerf}, \citenamefont {Wenger}, \citenamefont {Tualle-Brouri},\ and\ \citenamefont {Grangier}}]{10.5555/2011564.2011570}%
  \BibitemOpen
  \bibfield  {author} {\bibinfo {author} {\bibfnamefont {F.}~\bibnamefont {Grosshans}}, \bibinfo {author} {\bibfnamefont {N.~J.}\ \bibnamefont {Cerf}}, \bibinfo {author} {\bibfnamefont {J.}~\bibnamefont {Wenger}}, \bibinfo {author} {\bibfnamefont {R.}~\bibnamefont {Tualle-Brouri}},\ and\ \bibinfo {author} {\bibfnamefont {P.}~\bibnamefont {Grangier}},\ }\bibfield  {title} {\bibinfo {title} {Virtual entanglement and reconciliation protocols for quantum cryptography with continuous variables},\ }\href@noop {} {\bibfield  {journal} {\bibinfo  {journal} {Quantum Info. Comput.}\ }\textbf {\bibinfo {volume} {3}},\ \bibinfo {pages} {535–552} (\bibinfo {year} {2003}{\natexlab{b}})}\BibitemShut {NoStop}%
\bibitem [{\citenamefont {Louisell}\ \emph {et~al.}(1961)\citenamefont {Louisell}, \citenamefont {Yariv},\ and\ \citenamefont {Siegman}}]{PhysRev.124.1646}%
  \BibitemOpen
  \bibfield  {author} {\bibinfo {author} {\bibfnamefont {W.~H.}\ \bibnamefont {Louisell}}, \bibinfo {author} {\bibfnamefont {A.}~\bibnamefont {Yariv}},\ and\ \bibinfo {author} {\bibfnamefont {A.~E.}\ \bibnamefont {Siegman}},\ }\bibfield  {title} {\bibinfo {title} {Quantum fluctuations and noise in parametric processes. i.},\ }\href {https://doi.org/10.1103/PhysRev.124.1646} {\bibfield  {journal} {\bibinfo  {journal} {Phys. Rev.}\ }\textbf {\bibinfo {volume} {124}},\ \bibinfo {pages} {1646} (\bibinfo {year} {1961})}\BibitemShut {NoStop}%
\bibitem [{\citenamefont {Burnham}\ and\ \citenamefont {Weinberg}(1970)}]{PhysRevLett.25.84}%
  \BibitemOpen
  \bibfield  {author} {\bibinfo {author} {\bibfnamefont {D.~C.}\ \bibnamefont {Burnham}}\ and\ \bibinfo {author} {\bibfnamefont {D.~L.}\ \bibnamefont {Weinberg}},\ }\bibfield  {title} {\bibinfo {title} {Observation of simultaneity in parametric production of optical photon pairs},\ }\href {https://doi.org/10.1103/PhysRevLett.25.84} {\bibfield  {journal} {\bibinfo  {journal} {Phys. Rev. Lett.}\ }\textbf {\bibinfo {volume} {25}},\ \bibinfo {pages} {84} (\bibinfo {year} {1970})}\BibitemShut {NoStop}%
\bibitem [{\citenamefont {Usenko}\ and\ \citenamefont {Filip}(2016)}]{e18010020}%
  \BibitemOpen
  \bibfield  {author} {\bibinfo {author} {\bibfnamefont {V.~C.}\ \bibnamefont {Usenko}}\ and\ \bibinfo {author} {\bibfnamefont {R.}~\bibnamefont {Filip}},\ }\bibfield  {title} {\bibinfo {title} {Trusted noise in continuous-variable quantum key distribution: A threat and a defense},\ }\bibfield  {journal} {\bibinfo  {journal} {Entropy}\ }\textbf {\bibinfo {volume} {18}},\ \href {https://doi.org/10.3390/e18010020} {10.3390/e18010020} (\bibinfo {year} {2016})\BibitemShut {NoStop}%
\bibitem [{\citenamefont {Tserkis}\ \emph {et~al.}(2020)\citenamefont {Tserkis}, \citenamefont {Hosseinidehaj}, \citenamefont {Walk},\ and\ \citenamefont {Ralph}}]{PhysRevResearch.2.013208}%
  \BibitemOpen
  \bibfield  {author} {\bibinfo {author} {\bibfnamefont {S.}~\bibnamefont {Tserkis}}, \bibinfo {author} {\bibfnamefont {N.}~\bibnamefont {Hosseinidehaj}}, \bibinfo {author} {\bibfnamefont {N.}~\bibnamefont {Walk}},\ and\ \bibinfo {author} {\bibfnamefont {T.~C.}\ \bibnamefont {Ralph}},\ }\bibfield  {title} {\bibinfo {title} {Teleportation-based collective attacks in gaussian quantum key distribution},\ }\href {https://doi.org/10.1103/PhysRevResearch.2.013208} {\bibfield  {journal} {\bibinfo  {journal} {Phys. Rev. Res.}\ }\textbf {\bibinfo {volume} {2}},\ \bibinfo {pages} {013208} (\bibinfo {year} {2020})}\BibitemShut {NoStop}%
\bibitem [{\citenamefont {Bunandar}\ \emph {et~al.}(2015)\citenamefont {Bunandar}, \citenamefont {Zhang}, \citenamefont {Shapiro},\ and\ \citenamefont {Englund}}]{PhysRevA.91.022336}%
  \BibitemOpen
  \bibfield  {author} {\bibinfo {author} {\bibfnamefont {D.}~\bibnamefont {Bunandar}}, \bibinfo {author} {\bibfnamefont {Z.}~\bibnamefont {Zhang}}, \bibinfo {author} {\bibfnamefont {J.~H.}\ \bibnamefont {Shapiro}},\ and\ \bibinfo {author} {\bibfnamefont {D.~R.}\ \bibnamefont {Englund}},\ }\bibfield  {title} {\bibinfo {title} {Practical high-dimensional quantum key distribution with decoy states},\ }\href {https://doi.org/10.1103/PhysRevA.91.022336} {\bibfield  {journal} {\bibinfo  {journal} {Phys. Rev. A}\ }\textbf {\bibinfo {volume} {91}},\ \bibinfo {pages} {022336} (\bibinfo {year} {2015})}\BibitemShut {NoStop}%
\bibitem [{\citenamefont {Liu}\ \emph {et~al.}(2021)\citenamefont {Liu}, \citenamefont {Liu}, \citenamefont {Zhang},\ and\ \citenamefont {Huang}}]{LIU2021100119}%
  \BibitemOpen
  \bibfield  {author} {\bibinfo {author} {\bibfnamefont {J.-Y.}\ \bibnamefont {Liu}}, \bibinfo {author} {\bibfnamefont {X.}~\bibnamefont {Liu}}, \bibinfo {author} {\bibfnamefont {W.}~\bibnamefont {Zhang}},\ and\ \bibinfo {author} {\bibfnamefont {Y.-D.}\ \bibnamefont {Huang}},\ }\bibfield  {title} {\bibinfo {title} {Impact of fiber dispersion on the performance of entanglement-based dispersive optics quantum key distribution},\ }\href {https://doi.org/https://doi.org/10.1016/j.jnlest.2021.100119} {\bibfield  {journal} {\bibinfo  {journal} {Journal of Electronic Science and Technology}\ }\textbf {\bibinfo {volume} {19}},\ \bibinfo {pages} {100119} (\bibinfo {year} {2021})}\BibitemShut {NoStop}%
\bibitem [{\citenamefont {Wolf}\ \emph {et~al.}(2006)\citenamefont {Wolf}, \citenamefont {Giedke},\ and\ \citenamefont {Cirac}}]{PhysRevLett.96.080502}%
  \BibitemOpen
  \bibfield  {author} {\bibinfo {author} {\bibfnamefont {M.~M.}\ \bibnamefont {Wolf}}, \bibinfo {author} {\bibfnamefont {G.}~\bibnamefont {Giedke}},\ and\ \bibinfo {author} {\bibfnamefont {J.~I.}\ \bibnamefont {Cirac}},\ }\bibfield  {title} {\bibinfo {title} {Extremality of gaussian quantum states},\ }\href {https://doi.org/10.1103/PhysRevLett.96.080502} {\bibfield  {journal} {\bibinfo  {journal} {Phys. Rev. Lett.}\ }\textbf {\bibinfo {volume} {96}},\ \bibinfo {pages} {080502} (\bibinfo {year} {2006})}\BibitemShut {NoStop}%
\bibitem [{\citenamefont {Weedbrook}\ \emph {et~al.}(2012)\citenamefont {Weedbrook}, \citenamefont {Pirandola}, \citenamefont {Garc\'{\i}a-Patr\'on}, \citenamefont {Cerf}, \citenamefont {Ralph}, \citenamefont {Shapiro},\ and\ \citenamefont {Lloyd}}]{RevModPhys.84.621}%
  \BibitemOpen
  \bibfield  {author} {\bibinfo {author} {\bibfnamefont {C.}~\bibnamefont {Weedbrook}}, \bibinfo {author} {\bibfnamefont {S.}~\bibnamefont {Pirandola}}, \bibinfo {author} {\bibfnamefont {R.}~\bibnamefont {Garc\'{\i}a-Patr\'on}}, \bibinfo {author} {\bibfnamefont {N.~J.}\ \bibnamefont {Cerf}}, \bibinfo {author} {\bibfnamefont {T.~C.}\ \bibnamefont {Ralph}}, \bibinfo {author} {\bibfnamefont {J.~H.}\ \bibnamefont {Shapiro}},\ and\ \bibinfo {author} {\bibfnamefont {S.}~\bibnamefont {Lloyd}},\ }\bibfield  {title} {\bibinfo {title} {Gaussian quantum information},\ }\href {https://doi.org/10.1103/RevModPhys.84.621} {\bibfield  {journal} {\bibinfo  {journal} {Rev. Mod. Phys.}\ }\textbf {\bibinfo {volume} {84}},\ \bibinfo {pages} {621} (\bibinfo {year} {2012})}\BibitemShut {NoStop}%
\bibitem [{\citenamefont {Garc\'{\i}a-Patr\'on}\ and\ \citenamefont {Cerf}(2006)}]{Cerf_gaussian_optimality}%
  \BibitemOpen
  \bibfield  {author} {\bibinfo {author} {\bibfnamefont {R.}~\bibnamefont {Garc\'{\i}a-Patr\'on}}\ and\ \bibinfo {author} {\bibfnamefont {N.~J.}\ \bibnamefont {Cerf}},\ }\bibfield  {title} {\bibinfo {title} {Unconditional optimality of gaussian attacks against continuous-variable quantum key distribution},\ }\href {https://doi.org/10.1103/PhysRevLett.97.190503} {\bibfield  {journal} {\bibinfo  {journal} {Phys. Rev. Lett.}\ }\textbf {\bibinfo {volume} {97}},\ \bibinfo {pages} {190503} (\bibinfo {year} {2006})}\BibitemShut {NoStop}%
\bibitem [{\citenamefont {Navascu\'es}\ \emph {et~al.}(2006)\citenamefont {Navascu\'es}, \citenamefont {Grosshans},\ and\ \citenamefont {Ac\'{\i}n}}]{PhysRevLett.97.190502}%
  \BibitemOpen
  \bibfield  {author} {\bibinfo {author} {\bibfnamefont {M.}~\bibnamefont {Navascu\'es}}, \bibinfo {author} {\bibfnamefont {F.}~\bibnamefont {Grosshans}},\ and\ \bibinfo {author} {\bibfnamefont {A.}~\bibnamefont {Ac\'{\i}n}},\ }\bibfield  {title} {\bibinfo {title} {Optimality of gaussian attacks in continuous-variable quantum cryptography},\ }\href {https://doi.org/10.1103/PhysRevLett.97.190502} {\bibfield  {journal} {\bibinfo  {journal} {Phys. Rev. Lett.}\ }\textbf {\bibinfo {volume} {97}},\ \bibinfo {pages} {190502} (\bibinfo {year} {2006})}\BibitemShut {NoStop}%
\bibitem [{\citenamefont {Leverrier}\ and\ \citenamefont {Grangier}(2010)}]{PhysRevA.81.062314}%
  \BibitemOpen
  \bibfield  {author} {\bibinfo {author} {\bibfnamefont {A.}~\bibnamefont {Leverrier}}\ and\ \bibinfo {author} {\bibfnamefont {P.}~\bibnamefont {Grangier}},\ }\bibfield  {title} {\bibinfo {title} {Simple proof that gaussian attacks are optimal among collective attacks against continuous-variable quantum key distribution with a gaussian modulation},\ }\href {https://doi.org/10.1103/PhysRevA.81.062314} {\bibfield  {journal} {\bibinfo  {journal} {Phys. Rev. A}\ }\textbf {\bibinfo {volume} {81}},\ \bibinfo {pages} {062314} (\bibinfo {year} {2010})}\BibitemShut {NoStop}%
\bibitem [{\citenamefont {Renner}\ and\ \citenamefont {Cirac}(2009)}]{PhysRevLett.102.110504}%
  \BibitemOpen
  \bibfield  {author} {\bibinfo {author} {\bibfnamefont {R.}~\bibnamefont {Renner}}\ and\ \bibinfo {author} {\bibfnamefont {J.~I.}\ \bibnamefont {Cirac}},\ }\bibfield  {title} {\bibinfo {title} {de finetti representation theorem for infinite-dimensional quantum systems and applications to quantum cryptography},\ }\href {https://doi.org/10.1103/PhysRevLett.102.110504} {\bibfield  {journal} {\bibinfo  {journal} {Phys. Rev. Lett.}\ }\textbf {\bibinfo {volume} {102}},\ \bibinfo {pages} {110504} (\bibinfo {year} {2009})}\BibitemShut {NoStop}%
\bibitem [{\citenamefont {Pirandola}\ \emph {et~al.}(2008)\citenamefont {Pirandola}, \citenamefont {Braunstein},\ and\ \citenamefont {Lloyd}}]{PhysRevLett.101.200504}%
  \BibitemOpen
  \bibfield  {author} {\bibinfo {author} {\bibfnamefont {S.}~\bibnamefont {Pirandola}}, \bibinfo {author} {\bibfnamefont {S.~L.}\ \bibnamefont {Braunstein}},\ and\ \bibinfo {author} {\bibfnamefont {S.}~\bibnamefont {Lloyd}},\ }\bibfield  {title} {\bibinfo {title} {Characterization of collective gaussian attacks and security of coherent-state quantum cryptography},\ }\href {https://doi.org/10.1103/PhysRevLett.101.200504} {\bibfield  {journal} {\bibinfo  {journal} {Phys. Rev. Lett.}\ }\textbf {\bibinfo {volume} {101}},\ \bibinfo {pages} {200504} (\bibinfo {year} {2008})}\BibitemShut {NoStop}%
\bibitem [{\citenamefont {Renner}\ \emph {et~al.}(2005)\citenamefont {Renner}, \citenamefont {Gisin},\ and\ \citenamefont {Kraus}}]{PhysRevA.72.012332}%
  \BibitemOpen
  \bibfield  {author} {\bibinfo {author} {\bibfnamefont {R.}~\bibnamefont {Renner}}, \bibinfo {author} {\bibfnamefont {N.}~\bibnamefont {Gisin}},\ and\ \bibinfo {author} {\bibfnamefont {B.}~\bibnamefont {Kraus}},\ }\bibfield  {title} {\bibinfo {title} {Information-theoretic security proof for quantum-key-distribution protocols},\ }\href {https://doi.org/10.1103/PhysRevA.72.012332} {\bibfield  {journal} {\bibinfo  {journal} {Phys. Rev. A}\ }\textbf {\bibinfo {volume} {72}},\ \bibinfo {pages} {012332} (\bibinfo {year} {2005})}\BibitemShut {NoStop}%
\bibitem [{\citenamefont {Devetak}\ and\ \citenamefont {Winter}(2005)}]{Devetak_2005}%
  \BibitemOpen
  \bibfield  {author} {\bibinfo {author} {\bibfnamefont {I.}~\bibnamefont {Devetak}}\ and\ \bibinfo {author} {\bibfnamefont {A.}~\bibnamefont {Winter}},\ }\bibfield  {title} {\bibinfo {title} {Distillation of secret key and entanglement from quantum states},\ }\href {https://doi.org/10.1098/rspa.2004.1372} {\bibfield  {journal} {\bibinfo  {journal} {Proceedings of the Royal Society A: Mathematical, Physical and Engineering Sciences}\ }\textbf {\bibinfo {volume} {461}},\ \bibinfo {pages} {207–235} (\bibinfo {year} {2005})}\BibitemShut {NoStop}%
\bibitem [{\citenamefont {Cerf}\ and\ \citenamefont {Adami}(1996)}]{cerf1996accessible}%
  \BibitemOpen
  \bibfield  {author} {\bibinfo {author} {\bibfnamefont {N.~J.}\ \bibnamefont {Cerf}}\ and\ \bibinfo {author} {\bibfnamefont {C.}~\bibnamefont {Adami}},\ }\href@noop {} {\bibinfo {title} {Accessible information in quantum measurement}} (\bibinfo {year} {1996}),\ \Eprint {https://arxiv.org/abs/quant-ph/9611032} {arXiv:quant-ph/9611032 [quant-ph]} \BibitemShut {NoStop}%
\bibitem [{\citenamefont {Ali-Khan}\ \emph {et~al.}(2007)\citenamefont {Ali-Khan}, \citenamefont {Broadbent},\ and\ \citenamefont {Howell}}]{PhysRevLett.98.060503}%
  \BibitemOpen
  \bibfield  {author} {\bibinfo {author} {\bibfnamefont {I.}~\bibnamefont {Ali-Khan}}, \bibinfo {author} {\bibfnamefont {C.~J.}\ \bibnamefont {Broadbent}},\ and\ \bibinfo {author} {\bibfnamefont {J.~C.}\ \bibnamefont {Howell}},\ }\bibfield  {title} {\bibinfo {title} {Large-alphabet quantum key distribution using energy-time entangled bipartite states},\ }\href {https://doi.org/10.1103/PhysRevLett.98.060503} {\bibfield  {journal} {\bibinfo  {journal} {Phys. Rev. Lett.}\ }\textbf {\bibinfo {volume} {98}},\ \bibinfo {pages} {060503} (\bibinfo {year} {2007})}\BibitemShut {NoStop}%
\bibitem [{\citenamefont {Franson}(1992)}]{PhysRevA.45.3126}%
  \BibitemOpen
  \bibfield  {author} {\bibinfo {author} {\bibfnamefont {J.~D.}\ \bibnamefont {Franson}},\ }\bibfield  {title} {\bibinfo {title} {Nonlocal cancellation of dispersion},\ }\href {https://doi.org/10.1103/PhysRevA.45.3126} {\bibfield  {journal} {\bibinfo  {journal} {Phys. Rev. A}\ }\textbf {\bibinfo {volume} {45}},\ \bibinfo {pages} {3126} (\bibinfo {year} {1992})}\BibitemShut {NoStop}%
\bibitem [{\citenamefont {Gel'fand}\ and\ \citenamefont {Yaglom}(1959)}]{gel'fand1959calculation}%
  \BibitemOpen
  \bibfield  {author} {\bibinfo {author} {\bibfnamefont {I.}~\bibnamefont {Gel'fand}}\ and\ \bibinfo {author} {\bibfnamefont {A.}~\bibnamefont {Yaglom}},\ }\href {https://books.google.com/books?id=o0bSYgEACAAJ} {\emph {\bibinfo {title} {Calculation of the Amount of Information about a Random Function Contained in Another Such Function}}},\ American Mathematical Society translations\ (\bibinfo  {publisher} {American Mathematical Society},\ \bibinfo {year} {1959})\BibitemShut {NoStop}%
\bibitem [{\citenamefont {Lee}\ \emph {et~al.}(2018)\citenamefont {Lee} \emph {et~al.}}]{lee2018high}%
  \BibitemOpen
  \bibfield  {author} {\bibinfo {author} {\bibfnamefont {C.}~\bibnamefont {Lee}} \emph {et~al.},\ }\emph {\bibinfo {title} {High-dimensional quantum communication over deployed fiber}},\ \href@noop {} {Ph.D. thesis},\ \bibinfo  {school} {Massachusetts Institute of Technology} (\bibinfo {year} {2018})\BibitemShut {NoStop}%
\bibitem [{\citenamefont {Korzh}\ \emph {et~al.}(2020)\citenamefont {Korzh}, \citenamefont {Zhao}, \citenamefont {Allmaras}, \citenamefont {Frasca}, \citenamefont {Autry}, \citenamefont {Bersin}, \citenamefont {Beyer}, \citenamefont {Briggs}, \citenamefont {Bumble}, \citenamefont {Colangelo}, \citenamefont {Crouch}, \citenamefont {Dane}, \citenamefont {Gerrits}, \citenamefont {Lita}, \citenamefont {Marsili}, \citenamefont {Moody}, \citenamefont {Pe{\~{n}}a}, \citenamefont {Ramirez}, \citenamefont {Rezac}, \citenamefont {Sinclair}, \citenamefont {Stevens}, \citenamefont {Velasco}, \citenamefont {Verma}, \citenamefont {Wollman}, \citenamefont {Xie}, \citenamefont {Zhu}, \citenamefont {Hale}, \citenamefont {Spiropulu}, \citenamefont {Silverman}, \citenamefont {Mirin}, \citenamefont {Nam}, \citenamefont {Kozorezov}, \citenamefont {Shaw},\ and\ \citenamefont {Berggren}}]{Korzh2020}%
  \BibitemOpen
  \bibfield  {author} {\bibinfo {author} {\bibfnamefont {B.}~\bibnamefont {Korzh}}, \bibinfo {author} {\bibfnamefont {Q.-Y.}\ \bibnamefont {Zhao}}, \bibinfo {author} {\bibfnamefont {J.~P.}\ \bibnamefont {Allmaras}}, \bibinfo {author} {\bibfnamefont {S.}~\bibnamefont {Frasca}}, \bibinfo {author} {\bibfnamefont {T.~M.}\ \bibnamefont {Autry}}, \bibinfo {author} {\bibfnamefont {E.~A.}\ \bibnamefont {Bersin}}, \bibinfo {author} {\bibfnamefont {A.~D.}\ \bibnamefont {Beyer}}, \bibinfo {author} {\bibfnamefont {R.~M.}\ \bibnamefont {Briggs}}, \bibinfo {author} {\bibfnamefont {B.}~\bibnamefont {Bumble}}, \bibinfo {author} {\bibfnamefont {M.}~\bibnamefont {Colangelo}}, \bibinfo {author} {\bibfnamefont {G.~M.}\ \bibnamefont {Crouch}}, \bibinfo {author} {\bibfnamefont {A.~E.}\ \bibnamefont {Dane}}, \bibinfo {author} {\bibfnamefont {T.}~\bibnamefont {Gerrits}}, \bibinfo {author} {\bibfnamefont {A.~E.}\ \bibnamefont {Lita}}, \bibinfo {author} {\bibfnamefont {F.}~\bibnamefont {Marsili}}, \bibinfo {author} {\bibfnamefont
  {G.}~\bibnamefont {Moody}}, \bibinfo {author} {\bibfnamefont {C.}~\bibnamefont {Pe{\~{n}}a}}, \bibinfo {author} {\bibfnamefont {E.}~\bibnamefont {Ramirez}}, \bibinfo {author} {\bibfnamefont {J.~D.}\ \bibnamefont {Rezac}}, \bibinfo {author} {\bibfnamefont {N.}~\bibnamefont {Sinclair}}, \bibinfo {author} {\bibfnamefont {M.~J.}\ \bibnamefont {Stevens}}, \bibinfo {author} {\bibfnamefont {A.~E.}\ \bibnamefont {Velasco}}, \bibinfo {author} {\bibfnamefont {V.~B.}\ \bibnamefont {Verma}}, \bibinfo {author} {\bibfnamefont {E.~E.}\ \bibnamefont {Wollman}}, \bibinfo {author} {\bibfnamefont {S.}~\bibnamefont {Xie}}, \bibinfo {author} {\bibfnamefont {D.}~\bibnamefont {Zhu}}, \bibinfo {author} {\bibfnamefont {P.~D.}\ \bibnamefont {Hale}}, \bibinfo {author} {\bibfnamefont {M.}~\bibnamefont {Spiropulu}}, \bibinfo {author} {\bibfnamefont {K.~L.}\ \bibnamefont {Silverman}}, \bibinfo {author} {\bibfnamefont {R.~P.}\ \bibnamefont {Mirin}}, \bibinfo {author} {\bibfnamefont {S.~W.}\ \bibnamefont {Nam}}, \bibinfo {author}
  {\bibfnamefont {A.~G.}\ \bibnamefont {Kozorezov}}, \bibinfo {author} {\bibfnamefont {M.~D.}\ \bibnamefont {Shaw}},\ and\ \bibinfo {author} {\bibfnamefont {K.~K.}\ \bibnamefont {Berggren}},\ }\bibfield  {title} {\bibinfo {title} {Demonstration of sub-3 ps temporal resolution with a superconducting nanowire single-photon detector},\ }\href {https://doi.org/10.1038/s41566-020-0589-x} {\bibfield  {journal} {\bibinfo  {journal} {Nature Photonics}\ }\textbf {\bibinfo {volume} {14}},\ \bibinfo {pages} {250} (\bibinfo {year} {2020})}\BibitemShut {NoStop}%
\bibitem [{\citenamefont {Colangelo}\ \emph {et~al.}(2023)\citenamefont {Colangelo}, \citenamefont {Korzh}, \citenamefont {Allmaras}, \citenamefont {Beyer}, \citenamefont {Mueller}, \citenamefont {Briggs}, \citenamefont {Bumble}, \citenamefont {Runyan}, \citenamefont {Stevens}, \citenamefont {McCaughan}, \citenamefont {Zhu}, \citenamefont {Smith}, \citenamefont {Becker}, \citenamefont {Narv\'aez}, \citenamefont {Bienfang}, \citenamefont {Frasca}, \citenamefont {Velasco}, \citenamefont {Ramirez}, \citenamefont {Walter}, \citenamefont {Schmidt}, \citenamefont {Wollman}, \citenamefont {Spiropulu}, \citenamefont {Mirin}, \citenamefont {Nam}, \citenamefont {Berggren},\ and\ \citenamefont {Shaw}}]{PhysRevApplied.19.044093}%
  \BibitemOpen
  \bibfield  {author} {\bibinfo {author} {\bibfnamefont {M.}~\bibnamefont {Colangelo}}, \bibinfo {author} {\bibfnamefont {B.}~\bibnamefont {Korzh}}, \bibinfo {author} {\bibfnamefont {J.~P.}\ \bibnamefont {Allmaras}}, \bibinfo {author} {\bibfnamefont {A.~D.}\ \bibnamefont {Beyer}}, \bibinfo {author} {\bibfnamefont {A.~S.}\ \bibnamefont {Mueller}}, \bibinfo {author} {\bibfnamefont {R.~M.}\ \bibnamefont {Briggs}}, \bibinfo {author} {\bibfnamefont {B.}~\bibnamefont {Bumble}}, \bibinfo {author} {\bibfnamefont {M.}~\bibnamefont {Runyan}}, \bibinfo {author} {\bibfnamefont {M.~J.}\ \bibnamefont {Stevens}}, \bibinfo {author} {\bibfnamefont {A.~N.}\ \bibnamefont {McCaughan}}, \bibinfo {author} {\bibfnamefont {D.}~\bibnamefont {Zhu}}, \bibinfo {author} {\bibfnamefont {S.}~\bibnamefont {Smith}}, \bibinfo {author} {\bibfnamefont {W.}~\bibnamefont {Becker}}, \bibinfo {author} {\bibfnamefont {L.}~\bibnamefont {Narv\'aez}}, \bibinfo {author} {\bibfnamefont {J.~C.}\ \bibnamefont {Bienfang}}, \bibinfo {author} {\bibfnamefont
  {S.}~\bibnamefont {Frasca}}, \bibinfo {author} {\bibfnamefont {A.~E.}\ \bibnamefont {Velasco}}, \bibinfo {author} {\bibfnamefont {E.~E.}\ \bibnamefont {Ramirez}}, \bibinfo {author} {\bibfnamefont {A.~B.}\ \bibnamefont {Walter}}, \bibinfo {author} {\bibfnamefont {E.}~\bibnamefont {Schmidt}}, \bibinfo {author} {\bibfnamefont {E.~E.}\ \bibnamefont {Wollman}}, \bibinfo {author} {\bibfnamefont {M.}~\bibnamefont {Spiropulu}}, \bibinfo {author} {\bibfnamefont {R.}~\bibnamefont {Mirin}}, \bibinfo {author} {\bibfnamefont {S.~W.}\ \bibnamefont {Nam}}, \bibinfo {author} {\bibfnamefont {K.~K.}\ \bibnamefont {Berggren}},\ and\ \bibinfo {author} {\bibfnamefont {M.~D.}\ \bibnamefont {Shaw}},\ }\bibfield  {title} {\bibinfo {title} {Impedance-matched differential superconducting nanowire detectors},\ }\href {https://doi.org/10.1103/PhysRevApplied.19.044093} {\bibfield  {journal} {\bibinfo  {journal} {Phys. Rev. Appl.}\ }\textbf {\bibinfo {volume} {19}},\ \bibinfo {pages} {044093} (\bibinfo {year} {2023})}\BibitemShut
  {NoStop}%
\end{thebibliography}%
\end{document}